\documentclass[journal, twoside]{IEEEtran}
\usepackage{amsmath}
\usepackage{amsfonts}
\usepackage{upgreek}
\usepackage{algorithmic}
\usepackage{algorithm}
\usepackage{array}
\usepackage{textcomp}
\usepackage{stfloats}
\usepackage{url}
\usepackage{verbatim}
\usepackage{graphicx}
\usepackage{cite}
\usepackage{float}
\usepackage{threeparttable}
\usepackage{tabularx,colortbl}
\usepackage{booktabs}
\usepackage{multirow}
\usepackage{balance}
\usepackage{tablefootnote}
\usepackage{hyperref}
\hypersetup{colorlinks=true, linkcolor=blue, anchorcolor=blue, citecolor=blue}

\renewcommand{\arraystretch}{1.3}

\usepackage{subfigure}

\graphicspath{{fig/}}

\begin{document}

\title{HFORD: Hybrid Forward Optimization and Reverse Design Method and Its Applications to On-Chip Millimeter-Wave Inductive Elements}
\author{
    Yuzhen Song,~\IEEEmembership{Graduate Student Member,~IEEE},
    Yifan Wang,~\IEEEmembership{Graduate Student Member,~IEEE},\\
    Guqiao Chen,~\IEEEmembership{Graduate Student Member,~IEEE},
    Hanyu Liu,~\IEEEmembership{Graduate Student Member,~IEEE},\\
    Qi Wu,~\IEEEmembership{Member,~IEEE},
    Guangyi Lu,~\IEEEmembership{Member,~IEEE}, 
    Haiming Wang,~\IEEEmembership{Member,~IEEE},\\
    and Wei Hong,~\IEEEmembership{Fellow,~IEEE}

\thanks{Manuscript received June 11, 2026; revised **** **, ****; accepted **** **, ****. Date of publication ***** **, ****; date of current **** **, ****. This work was supported in part by the National Key Research and Development Program of China under Grant 2023YFB4404803, and in part by the National Natural Science Foundation of China under Grants 62371120 and 62574040. (\emph{Corresponding authors: Haiming Wang and Qi Wu.})}
\thanks{Yuzhen Song, Yifan Wang, Guqiao Chen, and Hanyu Liu are with the State Key Laboratory of Millimeter Waves and the School of Information Science and Engineering, Southeast University, Nanjing 211189, China, and also with the R\&D Department, National Center of Technology Innovation for EDA, Nanjing 210031, China (e-mail: yuzhensong@seu.edu.cn; yifanwang@seu.edu.cn; chenguqiao@seu.edu.cn; hyliu2021@seu.edu.cn). }
\thanks{Guangyi Lu is with the School of Integrated Circuits, Southeast University, Nanjing 211189, China, and also with the R\&D Department, National Center of Technology Innovation for EDA, Nanjing 210031, China (e-mail: guangyilu@seu.edu.cn).}
\thanks{Qi Wu, Haiming Wang, and Wei Hong are with the State Key Laboratory of Millimeter Waves and the School of Information Science and Engineering, Southeast University, Nanjing 211189, China, and also with the Pervasive Communication Research Center, Purple Mountain Laboratories, Nanjing 211111, China (e-mail: qiwu@seu.edu.cn; hmwang@seu.edu.cn; weihong@seu.edu.cn). }
\thanks{Color versions of one or more of the figures in this paper are available online at \url{http://ieeexplore.ieee.org}.}
\thanks{Digital Object Identifier \qquad\qquad}
}
       
\markboth{IEEE Transactions on Computer-Aided Design of Integrated Circuits and Systems}%
{Song \MakeLowercase{\textit{et al.}}: HFORD: Hybrid Forward Optimization and Reverse Design Method for On-Chip mmWave Inductive Elements}


\maketitle

\begin{abstract}
On-chip inductive elements are pivotal in determining both the silicon footprint and performance of millimeter-wave (mmWave) integrated circuits. However, the layout-level synthesis of these passive devices is severely challenged by highly nonlinear geometry-to-performance mappings, computationally expensive full-wave electromagnetic simulations, topology-dependent design spaces, and the inherent non-uniqueness of inverse design. To overcome these bottlenecks, we propose a hybrid forward optimization and reverse design (HFORD) method for the target-to-layout synthesis of mmWave inductive elements. Utilizing a unified core to map device-level requirements to layout-level seeds, HFORD structures direct device targets and translates circuit specifications into a hierarchical synthesis flow. Specifically, sparse-fitting sampling is introduced to improve coverage across critical performance regions, while compact response-fitting coefficients significantly reduce training dimensionality. The HFORD core integrates a random forest for topology selection, a variational autoencoder for spectral feature generation, a mixture density network for probabilistic inverse mapping, and particle swarm optimization for latent space exploration. This integration improves the feasibility of the generated layout seeds under design rule check (DRC) constraints. Two design examples demonstrate that the proposed method accelerates the design cycle from hours to minutes compared to conventional optimization methods.
\end{abstract}


\section{Introduction}
\label{Sec_Intr}
\IEEEPARstart{O}{n-chip} inductive elements, such as spiral inductors and transformers \cite{meyer2000design}, are fundamental building blocks for millimeter-wave (mmWave) integrated circuits (ICs) in emerging 5G/6G and automotive radar applications. They occupy a substantial chip area and critically affect system performance, including gain, noise figure, and power efficiency. However, designing these devices at mmWave frequencies remains a challenge. Severe parasitic effects, such as skin effect and substrate coupling, make analytical formulas inaccurate and require full-wave electromagnetic (EM) simulations (e.g., HFSS, EMX). Since these simulations are computationally expensive, the traditional experience-driven and trial-and-error design process becomes time-consuming and inefficient.

To tackle these challenges, heuristic optimization algorithms such as the genetic algorithm (GA) \cite{Cui2003, Ozgun2003} and the particle swarm optimization (PSO) algorithm \cite{Ashinnmanesh2008, Nayeri2013} have been employed to automate the exploration of high-dimensional parameters. However, they rely on iterative EM solver calls and suffer from slow convergence and high computational costs. Machine learning (ML) surrogate models, such as deep neural networks (DNNs) and Gaussian processes \cite{Wu2020, Wu2024, Koziel2021, Prado2019, Jacobs2013, He2024}, have been introduced to replace time-intensive simulations with millisecond-level inference. Although surrogate-assisted optimization improves evaluation efficiency, most existing approaches still follow a forward modeling paradigm in which electromagnetic responses are predicted from predefined geometric variables and fixed topologies, and the final geometry still requires an external optimization loop.

Data-driven reverse engineering has emerged as a promising paradigm for generating structures directly from specifications \cite{Xiao2018, Zhou2023}. Early attempts using multilayer perceptrons (MLPs) to learn the backward mapping \cite{Zhang2018, Prado20192} suffer from inverse non-uniqueness because multiple geometries can correspond to identical S-parameter targets, and the standard regression loss tends to converge to the statistical mean of possible solutions. This averaging effect is detrimental in engineering design since the resulting geometries may violate discrete physical constraints, such as non-integer turns, or fail to reproduce the desired EM specifications. Tandem networks and generative adversarial networks (GANs) have been proposed to mitigate this issue \cite{Gupta2023, Zhang2024}, but tandem networks are difficult to train and computationally heavy, while GANs often suffer from mode collapse and training instability. Moreover, most existing works target planar antennas or metasurfaces. For mmWave on-chip passives, which require strict DRC compliance and accurate control, these black-box models still lack sufficient physical constraints and interpretability.

Recent architectural innovations have attempted to address these limitations. A multivalued neural network (MNN) resolves one-to-many mappings by simultaneously outputting $N$ geometric sets \cite{Zhang2018MNN}, but it introduces computational overhead in the post-selection process based on forward models and lacks probabilistic confidence. Non-uniqueness is mitigated by employing a transfer function-based network to filter the generated structure using the forward branches of poles and residues derived from target $|S_{11}|$ curves in \cite{Yuan.metasurfaces}. However, its dependence on detailed target curves limits its practicality for real-world constraint-driven design. A modified conditional variational autoencoder (VAE) with a loss of the forward model is introduced in \cite{MCVAE}. Although it helps ensure the desired electromagnetic response, it essentially collapses the multi-modal inverse problem into a deterministic search and does not explicitly model the posterior distribution of geometric parameters, limiting solution-space diversity.

RF inductive element synthesis methods have explored template-free layout generation, physics-augmented neural modeling, and multi-template transformer synthesis \cite{karahan2024nat, chae2024pulserf, he2025motif}. These methods improve structural flexibility or modeling accuracy, but practical mmWave passive synthesis must also handle different levels of design objectives. Some tasks are directly described by device level quantities, including inductance $L$, quality factor $Q$, coupling coefficient $k$, self resonance frequency $SRF$, and maximum available gain $G_{\max}$, whereas the design of the matching network is defined by circuit level responses such as $S_{11}$, $S_{21}$, bandwidth, and impedance transformation. In this work, circuit level objectives are handled through a hierarchical route, where network level specifications are first translated into intermediate device level requirements, and the resulting layout seed is then refined by circuit level post optimization.

In this work, a hybrid forward optimization and reverse design (HFORD) method is proposed for synthesizing mmWave on-chip inductive elements. The proposed method organizes passive synthesis as a target-to-layout flow. Device level requirements are directly processed by the HFORD core, while circuit level specifications are first translated into intermediate device level requirements through a task-dependent initialization step. The HFORD core combines random-forest-based topology selection, VAE-based spectral feature generation, MDN-based probabilistic inverse mapping, and PSO-based latent space optimization to obtain a layout level initial solution. For complex circuit level tasks, the initial solution can be further refined by a task-dependent post optimization procedure.

The main contributions of this work are summarized as follows:
\begin{enumerate}
    \item A hierarchical HFORD method for synthesizing mmWave on-chip inductive elements that unifies device and circuit-level specifications while separating target translation, reusable HFORD core generation, and circuit optimization for versatile passive synthesis.
    \item A performance-aware dataset construction method using sparse-fitting sampling that reduces full-wave EM simulation costs. Expanding data in the performance space improves the boundary region coverage essential for inverse design.
    \item A physics-informed parametric modeling strategy for transformers that compresses high-dimensional broadband electrical responses into low-dimensional features without losing critical physical signatures like resonance and high-frequency saturation.
\end{enumerate}

The remainder of this paper is organized as follows. Section \ref{Sec_Backg} introduces the problem formulation. Section \ref{Sec_HFORD} details the proposed methodology, including the sparse-fitting sampling strategy for efficient dataset construction and the hybrid optimization flow. Section \ref{Sec_Results} provides two practical examples of inductive elements that are used to verify the proposed HFORD method in Section \ref{Sec_HFORD}. Section \ref{Sec_Conc} concludes this paper.

\section{Problem Formulation}
\label{Sec_Backg}
The passive synthesis tasks in this work are specified at either the device or circuit level.  The former is described by inductive element characteristics, while the latter is defined by network level responses.

Direct optimization of circuit level objectives over layout variables usually leads to a large and strongly coupled search space due to layout parasitics, loading effects, and topology dependent EM behavior. Therefore, circuit level specifications are first translated into intermediate device level requirements through task-dependent initialization, analytical estimation, or schematic level optimization. Both types of tasks are then unified as a target-to-layout synthesis problem, which is conventionally formulated as
\begin{equation}
\begin{aligned}
\min_{\mathbf{x}}\;& J\!\left(\mathbf{x},\mathbf{y}(\mathbf{x},f)\right)\\
\text{s.t.}\;& \mathbf{h}_{\text{elec}}\!\left(\mathbf{y}(\mathbf{x},f)\right)\le \mathbf{0},\;
              \mathbf{g}_{\text{drc}}(\mathbf{x})\le \mathbf{0},\\
            & \mathbf{x}\in[\mathbf{x}_{\text{lb}},\mathbf{x}_{\text{ub}}],
\end{aligned}
\label{eq:classical}
\end{equation}
where $\mathbf{x}\in\mathbb{R}^{N}$ collects the tunable geometric variables, $\mathbf{y}(\mathbf{x},f)\in\mathcal{Y}$ is the EM response taking values in the response space $\mathcal{Y}\subseteq\mathbb{R}^{C\times T}$ with $C$ port channels and $T$ frequency points, and $\mathbf{h}_{\text{elec}}$, $\mathbf{g}_{\text{drc}}$ encode electrical and process constraints.
 \begin{figure}[!t]
    \centering
    \includegraphics[width=3.0in]{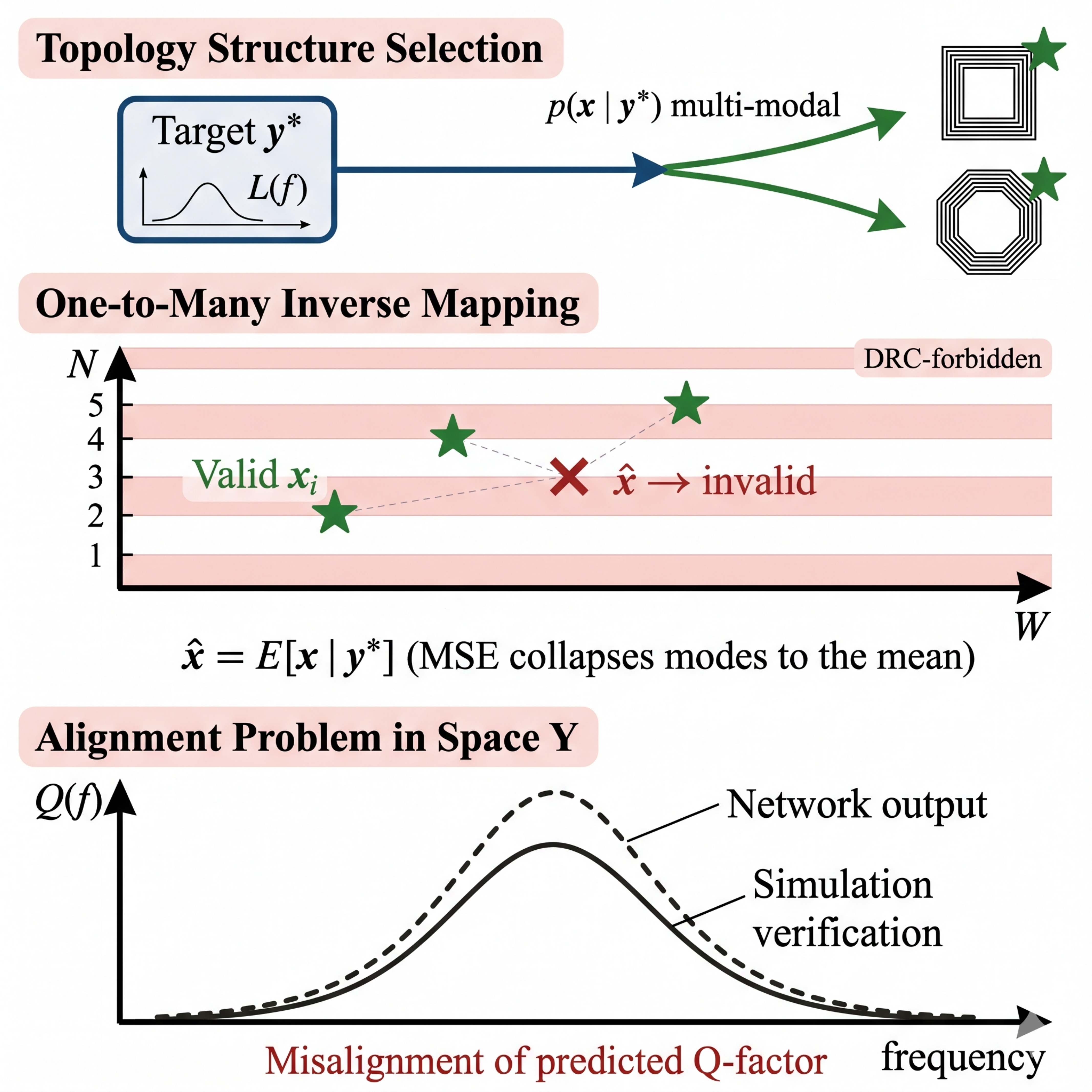}
    \caption{The core problem of inverse design.}
    \label{core problem}
    \vspace{-6pt}
\end{figure}
Equation~\eqref{eq:classical} implicitly assumes a fixed topology, a defined inverse map, and a faithful surrogate model. However, these assumptions generally break down at mmWave frequencies, as summarized in Fig.~\ref{core problem}, leading to three coupled ill-posed sub-problems.

\subsection{Topology Selection Problem}
\label{sec:problem-topology}
Let $\mathcal{T}=\{T_{1},\ldots,T_{K}\}$ be a library of admissible topologies. Each topology induces its own geometric space
$\Omega_{\text{phy}}^{(k)}$, the feasible region of the DRC $\Omega_{\text{drc}}^{(k)}$, and the forward EM operator $f_{k}:\Omega_{\text{phy}}^{(k)}\!\to\!\mathcal{Y}$. For a target $\bar{\mathbf{y}}$, the optimal topology is
\begin{equation}
\bar{T}=\arg\max_{T_{k}\in\mathcal{T}}\;
\max_{\mathbf{x}\in\Omega_{\text{phy}}^{(k)}\cap\,\Omega_{\text{drc}}^{(k)}}
\,-\bigl\|f_{k}(\mathbf{x})-\bar{\mathbf{y}}\bigr\|,
\label{eq:topology}
\end{equation}
which requires solving one inverse problem for each candidate topology. To reduce bias from manual topology selection, HFORD replaces \eqref{eq:topology} with a learned selector $\mathcal{C}:\bar{\mathbf{y}}\!\mapsto\!T_k$ that approximates $\bar{T}$ in a single inference.

\subsection{Inverse Non-uniqueness and Mode Collapse}
\label{sec:problem-inverse}
Once $T_{k}$ is fixed, the forward operator $f_{k}$ is generically non-unique, so the image is a disconnected and multi-modal set that is given by
\begin{equation}
\bar{X}(\bar{\mathbf{y}})=
\bigl\{\mathbf{x}\in\Omega_{\text{phy}}^{(k)}\cap\,\Omega_{\text{drc}}^{(k)}
\,\big|\,f_{k}(\mathbf{x})=\bar{\mathbf{y}}\bigr\}.
\label{eq:preimage}
\end{equation}
Any deterministic regressor $g_{\boldsymbol{\theta}}:\mathbf{y}\!\mapsto\!\mathbf{x}$ trained in MSE converges to the conditional mean
\begin{equation}
\bar{g}_{\boldsymbol{\theta}}(\bar{\mathbf{y}})=
\mathbb{E}\!\left[\mathbf{x}\mid\bar{\mathbf{y}}\right],
\label{eq:mse-mean}
\end{equation}
which may lie between feasible modes rather than on valid solutions. As shown in Fig.~\ref{core problem}, this averaged estimate can enter DRC forbidden regions, such as non-integer turns or subminimum line widths. The inverse task is therefore formulated as conditional density estimation with feasibility-aware mode selection, i.e., 
\begin{equation}
\bar{\mathbf{x}}\in
\arg\max_{\mathbf{x}\,\in\,\Omega_{\text{phy}}^{(k)}\cap\,\Omega_{\text{drc}}^{(k)}}
\;p(\mathbf{x}\mid\bar{\mathbf{y}}).
\label{eq:mode-select}
\end{equation}

\subsection{Optimization Illusion under Distributional Extrapolation}
The third pathology occurs after the surrogate-based optimizer has converged successfully. Let $\hat{f}_{\boldsymbol{\theta}}$ be a surrogate trained in $\mathcal{D}=\{(\mathbf{x}_{i},\mathbf{y}_{i})\}_{i=1}^{M}\sim p_{\mathcal{D}}$, and let $f$ be the ground truth solver. A typical loop returns
\begin{equation}
\begin{aligned}
\hat{\mathbf{x}}
&=
\arg\min_{\mathbf{x}}\,
J\!\bigl(\hat{f}_{\boldsymbol{\theta}}(\mathbf{x}),\bar{\mathbf{y}}\bigr),
\\
&\text{s.t.}\quad
J\!\bigl(\hat{f}_{\boldsymbol{\theta}}(\hat{\mathbf{x}}),\bar{\mathbf{y}}\bigr)
\approx 0 .
\end{aligned}
\label{eq:illusion}
\end{equation}
However, EM verification gives $J(f(\hat{\mathbf{x}}),\bar{\mathbf{y}})\!\gg\!0$ even with good held-out accuracy. This discrepancy is referred to here as the optimization illusion. It occurs when a global search minimizes the surrogate objective outside of the training support, especially near performance frontiers.

Generative surrogates further aggravate this issue. When latent optimization leaves the learned response manifold $\hat{M}_{\boldsymbol{\theta}}\subset\mathcal{Y}$, the decoder may still produce smooth but physically unrealizable spectra, such that $\hat{M}_{\boldsymbol{\theta}}\neq M_{f}=f(\Omega_{\text{phy}})$. As shown at the bottom of Fig.~\ref{core problem}, the predicted and simulated $Q(f)$ curves have similar shapes, while the constrained peak is shifted enough to miss the target.

\section{HFORD Method}
\label{Sec_HFORD}
This section introduces the proposed HFORD method, including response parameterization, dataset construction, offline model training, layout seed generation, and task-dependent post optimization.

\subsection{Offline and Online Hybrid Architecture}
\label{sec:HFORD}
\begin{figure*}[t]
\centering
\includegraphics[width=7.0 in]{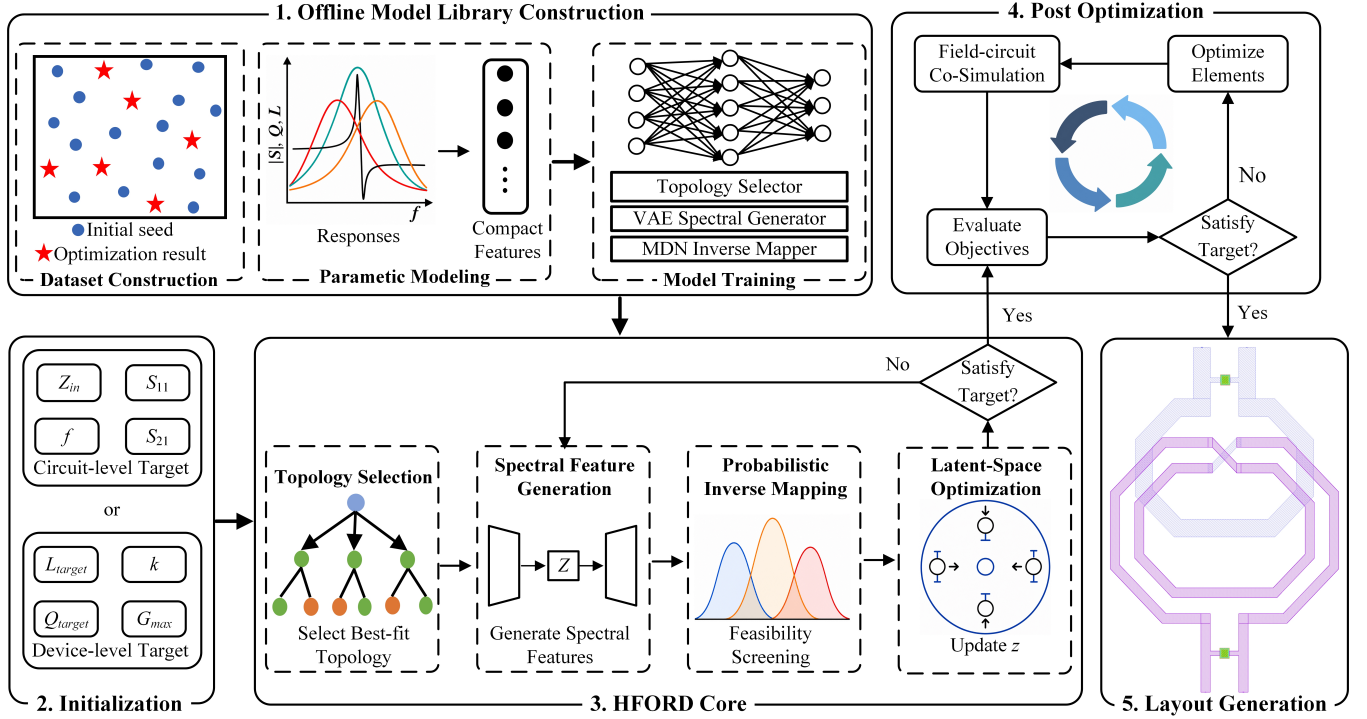}
\caption{Flowchart of the HFORD method.}
\label{HFORD}
\vspace{-6 pt}
\end{figure*}

The proposed HFORD method is organized as a hierarchical synthesis flow for mmWave on-chip inductive elements, as shown in Fig.~\ref{HFORD}. It consists of offline model library construction, online layout seed generation through the HFORD core, and task-dependent circuit level post optimization. The offline stage is completed before deployment and constructs reusable models. These models support online synthesis but do not enter the optimization loop for each new task.

With the offline library prepared, the online synthesis stage starts with a new design objective. For device level synthesis, the requirements are directly specified by inductive element quantities such as $L$, $Q$, $k$, $G_{\max}$, and DRC constraints. For circuit level synthesis, network level specifications such as $S_{11}$, $S_{21}$, input impedance, bandwidth, and insertion loss are first converted into intermediate device level requirements through task-dependent target translation, analytical modeling, schematic level optimization, or prior circuit knowledge. After this conversion, both types of tasks share a common device level requirement interface.

The HFORD core maps the unified device level requirements to a layout level seed. It selects a suitable topology, loads the corresponding pre-trained models, and performs layout seed generation in a continuous latent space. In this space, the candidate spectral features, the probabilistic inverse mapping results, and the constraints related to the DRC are jointly evaluated. The optimized latent variable is then converted into physical geometry parameters, producing an initial solution with the selected topology and device dimensions.

For device level synthesis, the generated seed can be verified by full-wave EM simulations and used as the final inductive element layout if the targets are satisfied. For circuit level synthesis, the same seed serves as a meaningful starting point for task-dependent post optimization, where the final objective is evaluated by complete field-circuit co-simulation rather than isolated device level characteristics.

Therefore, the proposed method separates offline model preparation, online seed generation, and task-dependent circuit level refinement. This modular organization allows the HFORD core to be reused in different passive synthesis tasks without being tied to a particular matching theory, compensation strategy, or circuit topology.

\subsection{Physics-Aware Parametric Modeling}
\label{Sec_Physics_Aware}
The accurate design of mmWave on-chip inductive elements requires full-band EM response modeling rather than single frequency predictions. However, directly using dense frequency samples as learning targets results in a high-dimensional output space, where small shifts in resonant frequency or quality factor peaks may cause large numerical errors despite similar physical trends.

For on-chip inductors, previous work \cite{wei2023highly} has shown that complex responses can be represented by key physical landmarks, providing a compact basis for the inductor examples in this work.

Direct black-box compression may lose important features such as resonance location, peak sharpness, high-frequency saturation, and coupling. To address this issue, a physics-informed parametric modeling strategy is introduced for transformer response representation. The broadband responses $L(f)$, $Q(f)$, $M(f)$, and $\operatorname{Re}\{Z_{21}\}$ are expressed by analytical fitting formulas, transforming dense frequency samples into compact features with physical meanings. This representation reduces learning dimensionality while preserving the resonance, loss, and coupling signatures.

First, $L(f)$ is modeled to capture low-frequency stability and a steep slope near the SRF as
\begin{equation}
    \label{Lcurve_eqt}
    L(f) = - \frac{\lambda_1}{(f - f_{r,1}) + \frac{\delta_1}{f - f_{r,1}}} - \frac{\lambda_2}{(f - f_{r,2}) + \frac{\delta_2}{f - f_{r,2}}} + L_{\text{dc}},
\end{equation}
where $L_{\text{dc}}$ denotes the baseline $L$, and the summation terms account for dynamic deviations from the primary and secondary self resonances. Here, $\lambda_i$ quantifies the resonant strength, $f_{r,i}$ the pole frequency, and $\delta_i$ a shape factor that regulates the trajectory of the pole.

Similarly, the positive region of $Q(f)$ is fitted with a distinct function that captures its peak magnitude and location
\begin{equation}
    \label{Qcurve_eqt}
    \begin{split}
    Q(f) &= \widetilde{Q}_{max1} \sin{\left( \frac{\pi f^{\widetilde{F}_{\rm Q_{max1}}}}{{\widetilde{F}_{\rm SRF1}}^{\widetilde{F}_{\rm Q_{max1}}}} \right)}^{\widetilde{Q}_{\rm sharp1}} \\
    & + \widetilde{Q}_{max2} \sin{\left( \frac{\pi (f - F_{set})^{\widetilde{F}_{\rm Q_{max2}}}}{({\widetilde{F}_{\rm SRF2}} - F_{set})^{\widetilde{F}_{\rm Q_{max2}}}} \right)}^{\widetilde{Q}_{\rm sharp2}},
    \end{split}
\end{equation}
where $\widetilde{Q}_{\rm max}$ represents the peak $Q$, $\widetilde{F}_{\rm Q_{max}}$ controls the peak frequency, and $\widetilde{Q}_{\rm sharp}$ determines the peak sharpness. $\widetilde{F}_{\rm SRF}$ denotes the pole frequency, and $F_{ set}$ denotes the valley frequency between the two fitted $Q$ peaks.

To separately capture magnetic energy storage and dissipation, the $Z_{21}$ is decomposed into its real part $\operatorname{Re}\{Z_{21}\}$ and the effective mutual inductance $M(f)$, both of which must be modeled to reconstruct $k(f)$.

First, $\operatorname{Re}\{Z_{21}\}$ captures AC loss dominated by skin effects and substrate coupling, modeled by a resonance-like rational function
\begin{equation}
\label{eq:Re_Z21}
\operatorname{Re}\{Z_{21}(f)\} = \frac{\alpha_{Z} f^{\beta_{Z}}}{(f - f_{c})^2 + \gamma_{Z}},
\end{equation}
where $\alpha_{Z}$ is the magnitude scaling factor, $f_{c}$ the center frequency of the resistive loss peak, $\beta_{Z}$ the peak sharpness, and $\gamma_{Z}$ the bandwidth and damping of the loss profile.

Second, the effective mutual inductance $M(f)$ deviates from its static value due to distributed capacitive parasitics, modeled by a composite function with DC bias and resonant terms
\begingroup
\small
\begin{equation}
\label{eq:M_fit_expanded}
M(f) =  - \frac{\eta_1}{(f - f_{m,1}) + \frac{\zeta_1}{f - f_{m,1}}} - \frac{\eta_2}{(f - f_{m,2}) + \frac{\zeta_2}{f - f_{m,2}}} + M_{\text{dc}},
\end{equation}
\endgroup
where $M_{\text{dc}}$ is the baseline mutual inductance, and the summation terms capture dynamic deviations from the primary and secondary resonant perturbations. Here, $\eta_i$ quantifies the resonant strength of $M(f)$, $f_{m,i}$ denotes the corresponding pole frequency, and $\zeta_i$ regulates the pole trajectory.

The proposed method is validated on 1,000 simulated transformer samples using EMX. As shown in Table~\ref{fitting_error_tab}, the fitted responses agree well with the EM data across the mmWave band and capture both low-frequency stability and near-resonance variations. The fitted coefficients form a compact physical feature vector, which is used as the conditional input of the probabilistic inverse mapper.
\begin{table}
    \centering
    \caption{Errors of Fitting Results}
    \label{fitting_error_tab}
    \begin{tabular}{|c|c|c|}
    \hline
    Structure & Interleaved & Symmetrical\\
    \hline
    $\operatorname{NRMSE}\{L\}$ & 1.72\% & 0.95\% \\
    \hline
    $\operatorname{NRMSE}\{Q\}$ & 2.24\% & 1.09\% \\
    \hline
    $\operatorname{NRMSE}\{M\}$ & 0.51\% & 0.53\% \\
    \hline
    $\operatorname{NRMSE}\{Z_{re}\}$ & 1.26\% & 1.12\% \\
    \hline
    \end{tabular}
    \vspace{-6 pt}
\end{table}

\subsection{Performance-Aware Dataset Construction}
\label{sec:dataset}
In the design of mmWave on-chip inductive elements, a single device category often spans a large design space, and exhaustive EM sampling becomes computationally expensive. Conventional samplers such as Latin hypercube sampling (LHS) \cite{McKay1979LHS} and Monte Carlo methods \cite{Forrester2008Surrogate} generate geometric parameters $x\in \mathcal{X}$ under predefined distributions. 

However, the nonlinear geometry-to-physics mapping makes the resulting performance samples $y\in \mathcal{Y}$ unevenly distributed, often clustered near common response regions while missing boundary regions.

To improve coverage of the performance space, an adaptive sparse fitting sampling strategy was adopted. As described in Algorithm~\ref{alg:sampling}, an initial dataset $\mathcal{D}$ is first generated by LHS and then expanded using a maximin criterion in the performance space. Each new sample is selected as
\begin{equation}
\mathbf{x}^* = \arg\max_{\mathbf{x} \in \mathcal{X}} \left( \min_{(\mathbf{x}_k, \mathbf{y}_k) \in \mathcal{D}} \| \hat{f}(\mathbf{x}) - \mathbf{y}_k \|_2 \right),
\end{equation}
where $\hat{f}(\cdot)$ denotes the surrogate model trained on the available samples. The selected candidates are then evaluated again by EM simulation before being added to $\mathcal{D}$, which ensures that all labels in the database are obtained from EM simulation. Since this objective is nonconvex and contains multiple sparse regions, the Niching migratory multi-swarm optimizer (NMMSO) \cite{NMMSO} is used to identify various candidate geometries in one run.
\begingroup
\makeatletter
\renewcommand{\fnum@algorithm}{Algorithm 1}
\makeatother

\begin{algorithm}[t]
    \caption{Performance-Aware Adaptive Sampling}
    \label{alg:sampling}
    \renewcommand{\algorithmicrequire}{\textbf{Input:}}
    \renewcommand{\algorithmicensure}{\textbf{Output:}}
    \begin{algorithmic}[1]
        \REQUIRE Geometric Sampling Boundary $\mathcal{X}$, Initial dataset size $N_{\mathrm{init}}$, Target Dataset Size $N_{target}$
        \ENSURE Dataset $\mathcal{D} = \{(x_i, y_i)\}_{i=1}^{N_{target}}$
        
        \STATE \textit{Step 1: Initialization}
        \STATE Initialize dataset $\mathcal{D}$ with $N_{init}$ samples using LHS
        \STATE Evaluate performance $y_i = f(x_i)$ for all initial samples
        
        \STATE \textit{Step 2: Adaptive Expansion Loop}
        \WHILE{$|\mathcal{D}| < N_{target}$}
            \STATE \textit{// Define the objective: Maximize distance to the nearest neighbor in Y-space}
            \STATE Define Objective Function $J(x)$:
            \STATE \quad $J(x) = \min_{(x_k, y_k) \in \mathcal{D}} \| \hat{f}(x) - y_k \|_2$
            
            \STATE \textit{Use NMMSO to find diverse candidates in X-space that maximize J(x)}
            \STATE \textit{NMMSO returns a set of local optima (candidates)}
            \STATE $X_{candidates} \leftarrow \text{NMMSO}(\arg\max_{x \in \mathcal{X}} J(x))$
            
            \FOR{each $x_{cand}$ in $X_{candidates}$}
                \IF{$|\mathcal{D}| < N_{target}$}
                    \STATE Evaluate $y_{cand} = f(x_{cand})$
                    \STATE $\mathcal{D} \leftarrow \mathcal{D} \cup \{(x_{cand}, y_{cand})\}$
                \ENDIF
            \ENDFOR
        \ENDWHILE
        
        \RETURN $\mathcal{D}$
    \end{algorithmic}
\end{algorithm}

\endgroup

In experiments, a 1:1 ratio is maintained between samples from the initial random sampling and those from the proposed adaptive optimization. This hybrid composition allows the model to retain global statistical coverage from random sampling while gaining boundary precision from the adaptive phase. This sampling strategy introduces additional computational overhead from the optimization loop in each iteration. When the EM simulations themselves are costly, this trade-off becomes justifiable. By selecting the most informative samples, our method reduces the redundant simulations required for target accuracy and improves data generation efficiency.

In addition to improving sample placement within a given topology, the efficient construction of a multiple topology model library also requires reducing the data burden across related structures. Although different topologies have different geometric ranges and response distributions, they share common EM mechanisms. Therefore, transfer learning is used to reuse spectral features learned from a source domain $\mathcal{D}_S$ for a related target domain $\mathcal{D}_T$.

The transfer process comprises source pre-training, parameter transfer, and fine-tuning. A baseline model is first trained on a representative structure, and its optimized weights $\boldsymbol{\theta}_S$ are then used to initialize the target model. The target model is finally fine-tuned using a small dataset, approximately 20\% of the source dataset, to adapt the mapping to the new topology. This strategy improves sample efficiency and enables rapid expansion of the offline component library without exhaustive EM simulations for every topology.

Quadrilateral and octagonal inductors in single-ended, symmetrical, and stacked configurations are evaluated. Using the single-ended and stacked quadrilateral inductors as base templates, parameter-transfer training is applied to the remaining types. The training comparison results are shown in Table~\ref{training_results}.
\begin{table}[htbp]
\caption{Training Results of Models for All Inductor Types.}
\label{training_results}
\centering
\begin{threeparttable}
\begin{tabular}{|c|c|c|c|}
\hline
Structure & Dataset & NRMSE1 & NRMSE2\\
\hline
Se-N-4  & 5000 & 3.6\% &   \\
\hline
Se-N-8  & 1000 & 4.8\% & 19.4\%  \\
\hline
S-N-4   & 1000 & 8.2\% & 15.9\% \\
\hline
S-N-8   & 1000 & 9.9\% & 17.5\% \\
\hline
Se-O-4  & 5000 & 6.5\% &  \\
\hline
Se-O-8  & 1000 & 9.3\% & 23.1\% \\
\hline
S-O-4   & 1000 & 11.3\% & 21.7\% \\
\hline
S-O-8   & 1000 & 9.7\% & 18.6\% \\
\hline
\end{tabular}
\begin{tablenotes}[flushleft]
\footnotesize
\item NRMSE1 obtained with the transfer learning model.
\item  NRMSE2 obtained without transfer learning under the same data volume.
\item Se/S denote single-ended/differential, N/O denote normal/overlapped, and 4/8 denote quadrilateral/octagonal layouts.
\end{tablenotes}
\end{threeparttable}
\vspace{-6 pt}
\end{table}

\subsection{Offline Model Training}
After dataset construction and response parameterization, the offline stage trains the reusable models required by the online HFORD core. For each topology, compact spectral features and geometric parameters are used to train the VAE-based spectral generator and the MDN-based probabilistic inverse mapper. A random forest selector is also trained across the device library to identify a suitable topology for each target requirement. Together, these pre-trained models form the offline model library for subsequent layout seed generation.

\subsubsection{Automated Device Type Selection}
\ 
\newline
\indent Given a target requirement vector, the first step of the online HFORD core is to select a suitable topology from the offline model library. This selection avoids searches in unsuitable design spaces and determines which VAE and MDN models are loaded for layout seed generation. In this work, a random-forest-based classifier is employed for automated device type selection.

Let $\mathcal{T}=\{T_1,T_2,\ldots,T_K\}$ denote the candidate topology library, and let $\mathbf{r}$ denote the target requirement vector, which includes $L_{WF}$, $Q_{WF}$, working frequency ($WF$), $SRF$, $k$, or $G_{\max}$ depending on the device type. 
A random forest consists of $N_t$ decision trees trained with bootstrap resampling and random feature selection \cite{breiman2001random}. Each decision tree outputs a candidate topology label $h_i(\mathbf{r})\in\mathcal{T}$. The final selected topology is determined by majority voting as
\begin{equation}
C(\mathbf{r})=
\arg\max_{T_k\in\mathcal{T}}
\sum_{i=1}^{N_t}
\mathbb{I}\left(h_i(\mathbf{r})=T_k\right),
\end{equation}
where $C(\mathbf{r})$ is the selected topology and $\mathbb{I}(\cdot)$ is the indicator function. The selected topology is then used to load the corresponding VAE and MDN models for latent space optimization.

\subsubsection{Generative Modeling of Spectral Features}
\ 
\newline
\indent The VAE \cite{Kingma2014Auto} is used to learn a compact and continuous latent representation $\mathbf{z}$ from the high-dimensional EM response 
$\mathbf{y}\in\mathbb{R}^{C\times T}$. The encoder maps the multi-channel spectral response $\mathbf{y}$ into a latent feature $\mathbf{z}$, while the decoder reconstructs the corresponding response $\hat{\mathbf{y}}$. This continuous latent space allows small perturbations in $\mathbf{z}$ to produce smooth variations in the reconstructed EM response, which facilitates subsequent latent space optimization.

Standard VAEs usually assume independent Gaussian likelihoods and therefore ignore correlations among different S-parameters and adjacent frequency points. To include these dependencies, the decoder likelihood is modeled with a Kronecker-structured covariance~\cite{8132183}. Let $\tilde{\mathbf{y}}=\mathrm{vec}(\mathbf{y})\in\mathbb{R}^{CT}$ denote the vectorized EM response, and let $\boldsymbol{\mu}_{\theta}(\mathbf{z}) =\mathrm{vec}(\hat{\mathbf{y}}_{\theta}(\mathbf{z}))$ denote the decoder mean. The likelihood is written as
\begin{equation}
p_{\theta}(\tilde{\mathbf{y}}|\mathbf{z}) =\mathcal{N}\left(\tilde{\mathbf{y}};\boldsymbol{\mu}_{\theta}(\mathbf{z}),\boldsymbol{\Sigma}_{c}\otimes\boldsymbol{\Sigma}_{f}\right),
\label{eq:vae_kron_likelihood}
\end{equation}
where $\boldsymbol{\Sigma}_{c}\in\mathbb{R}^{C\times C}$ captures coupling among different channels and
$\boldsymbol{\Sigma}_{f}\in\mathbb{R}^{T\times T}$ models smoothness across frequency points. This structure encourages physically consistent spectral reconstruction instead of treating all frequency samples independently.
\begin{figure}[tpb]
    \centering
    \subfigure[][]{%
    \label{L_vae}%
    \includegraphics[width=1.8in]{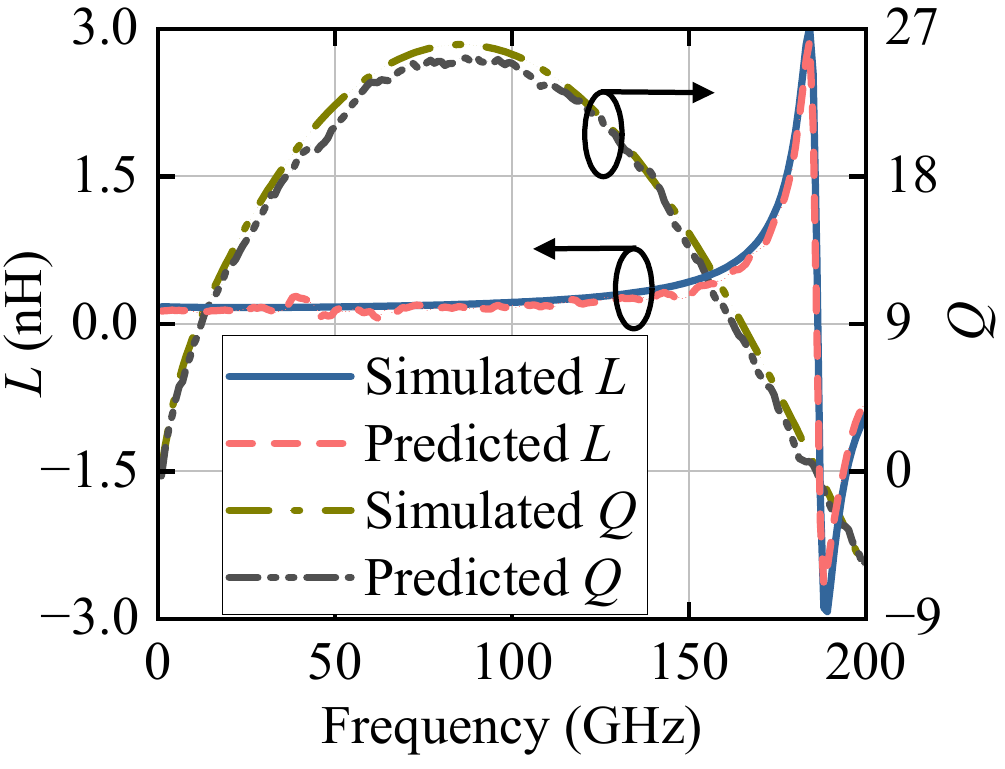}}
    \hspace{1pt}%
    \subfigure[][]{%
    \label{PCA_AE_VAE}%
    \includegraphics[width=1.62in]{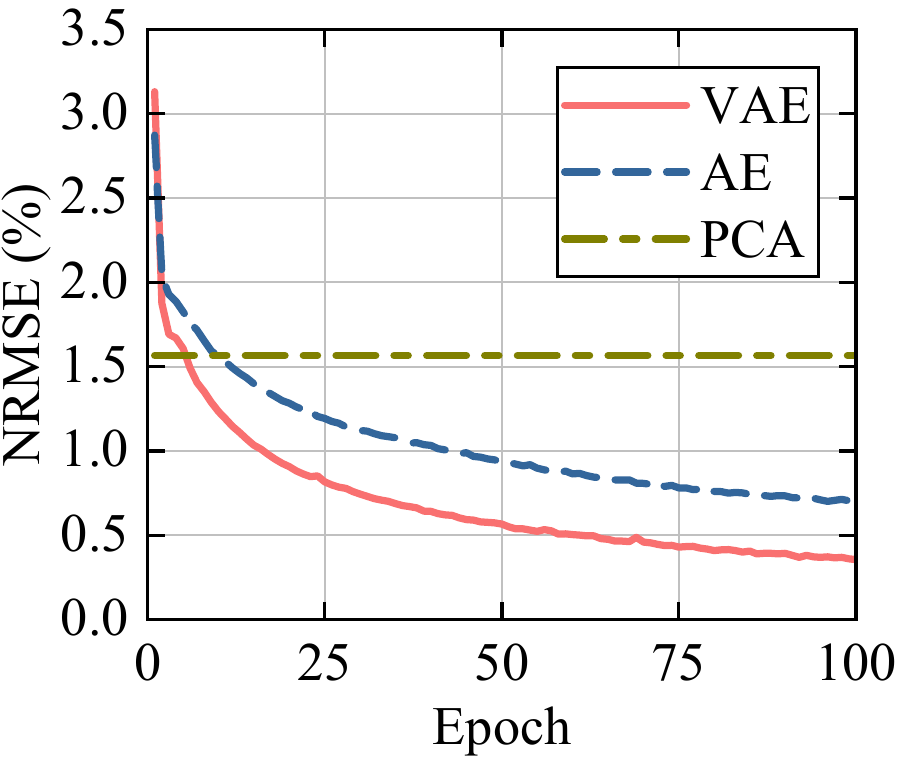}}
    \hspace{1pt}%
    \caption{Comparison between the simulated ground truth and the reconstructed curves predicted by the proposed VAE model. \subref{L_vae} $L$ and $Q$ reconstruction; \subref{PCA_AE_VAE} PCA, AE, and VAE.}%
    \label{fig_vae}
    \vspace{-6 pt}
\end{figure}
\begin{table}
    \centering
    \caption{Comparisons of the VAE with the PCA and the AE}
    \label{VAE_tab}
    \begin{tabular}{|c|c|c|c|}
    \hline
    Item & PCA & AE & VAE\\
    \hline
    Output dimension   &10&6&6\\
    \hline
    Model training time (s)  &1.2&38.5&46.1 \\
    \hline
    Prediction time ($\upmu$s) &21&11&10 \\
    \hline
    Mean NRMSE (\%) &1.65& 0.51&0.35 \\
    \hline
    \end{tabular}
    \vspace{-6 pt}
\end{table}

The encoder $q_\phi(\mathbf{z}|\mathbf{y})$ uses 1D convolutional layers and fully connected layers to parameterize the latent mean and variance, while the decoder reconstructs the full spectral response. The model is trained by maximizing the evidence lower bound, i.e.,
\begin{equation}
\begin{split}
\mathcal{L}(\theta, \phi; \mathbf{y}) &= 
\mathbb{E}_{q_\phi(\mathbf{z}|\mathbf{y})}[\log p_\theta(\mathbf{y}|\mathbf{z})] - D_\mathrm{KL}\big(q_\phi(\mathbf{z}|\mathbf{y}) \| p(\mathbf{z})\big),
\end{split}
\end{equation}
where $p(\mathbf{z}) = \mathcal{N}(\mathbf{0}, \mathbf{I})$.

After training, the proposed VAE is compared with principal component analysis (PCA) and a standard autoencoder (AE). As shown in Fig.~\ref{fig_vae}\subref{L_vae}, the reconstructed $L$ and $Q$ curves agree well with the simulated EM results. Fig.~\ref{fig_vae}\subref{PCA_AE_VAE} and Table~\ref{VAE_tab} further show that the VAE achieves the lowest mean NRMSE of $0.35\%$ with an output dimension of 6 and an online prediction time of $1.0\times10^{-5}$ s. Although the VAE requires a slightly longer offline training time, its online inference cost remains negligible for the proposed method.

\subsubsection{Probabilistic Inverse Mapping}
\ 
\newline
\indent The VAE decoder maps the latent space $\mathcal{Z}$ to the spectral space $\mathcal{Y}$, while design automation requires the inverse retrieval of geometric parameters $\mathbf{x}$ from a desired response. Since multiple geometries may satisfy the same spectral specification, a deterministic regressor trained with MSE tends to predict an averaged solution that may be physically invalid. To model this one-to-many mapping, an MDN \cite{Bishop2006} is used to predict the conditional distribution of geometric parameters $p(\mathbf{x}|\mathbf{z})$ from the VAE encode latent feature $\mathbf{z}$.

In standard MDN implementations, exponential activation is often used to enforce positive variance, but $\exp(a_i)$ may cause overflow or variance collapse when the unconstrained output $a_i$ becomes too large or too negative. To improve numerical stability, the standard deviation is predicted using a softplus activation,
\begin{equation}
\sigma_i = \ln(1+\exp(a_i))+\epsilon,
\end{equation}
where $a_i$ is the output of the unconstrained network and $\epsilon$ is a small positive constant. As shown in Table~\ref{MDN_tab}, this modification improves convergence stability. The standard exponential formulation converged in only 3 out of 10 trials, while the softplus formulation achieved a 100\% success rate with faster convergence in the deep training stage.


\begin{figure}[tpb]
\centering
\subfigure[][]{%
\label{L_mdn}%
\includegraphics[width=1.77in]{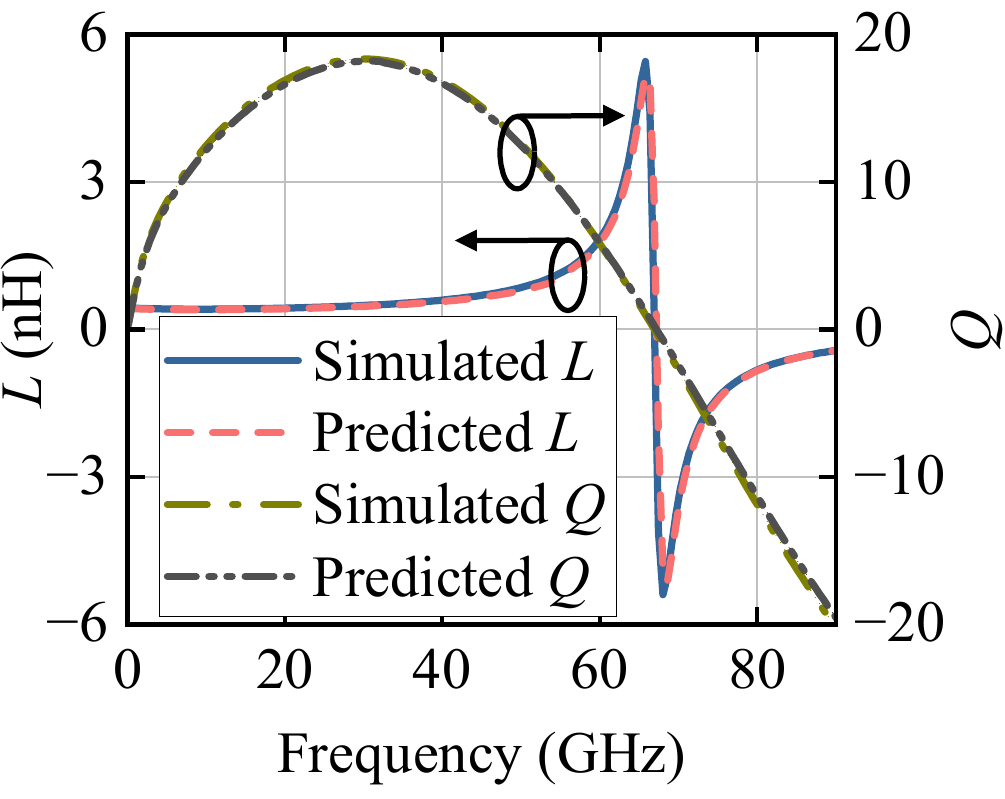}}
\hspace{-1pt}%
\subfigure[][]{%
\label{Q_mdn}%
\includegraphics[width=1.62in]{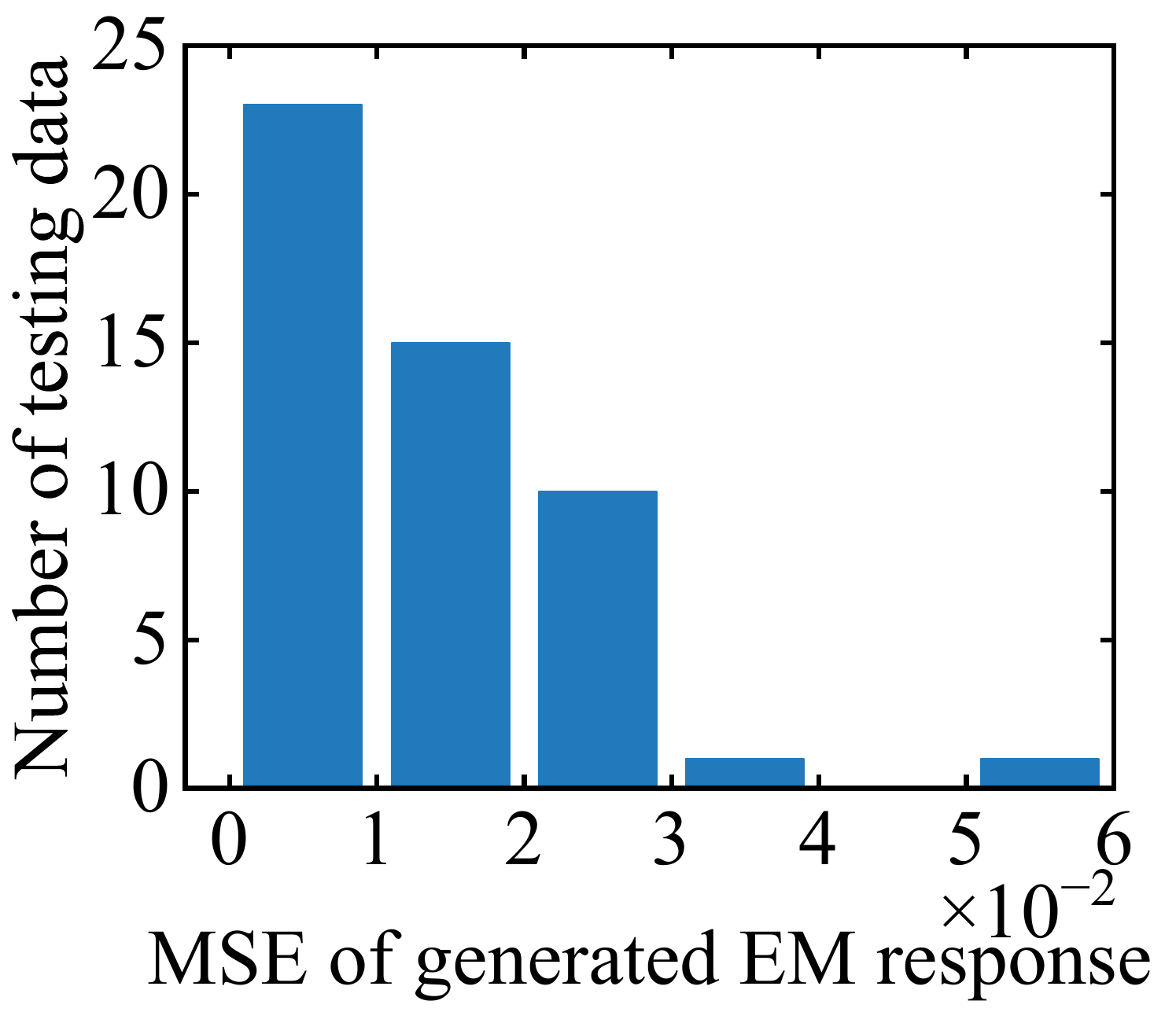}}
\hspace{1pt}%
\caption{Comparison between MDN prediction and the original data. \subref{L_mdn} $L$ and $Q$; \subref{Q_mdn} MSE.}%
\label{MDN_result}
\vspace{-6 pt}
\end{figure}

\subsection{HFORD Core for Layout Level Initial Solution Generation}
\label{sec:HFORD_core}
The HFORD core performs online layout level initial solution generation based on the unified device level requirements, as illustrated in Fig.~\ref{HFORD}. It integrates topology selection, VAE-based spectral generation, MDN-based probabilistic inverse mapping, and PSO-based latent space optimization. Instead of directly searching in the physical geometry space, which is often discontinuous due to topology dependent variables and DRC constraints, the HFORD core conducts the main optimization in the continuous latent space $\mathcal{Z}$.     

First, the workflow is initialized with the input target performance metrics of the passive device and the corresponding fabrication constraints. A pre-trained random forest classifier first selects the most suitable device topology according to the input targets and process limits. The pre-trained forward and inverse neural network models corresponding to the selected structure are then loaded. This step avoids unnecessary searches in unsuitable design spaces.
\begin{table}
    \centering
    \caption{Comparisons of the modified MDN with the MDN and the MLP}
    \label{MDN_tab}
    \begin{tabular}{|c|c|c|c|}
    \hline
    Item & MLP & MDN & Soft MDN \\
    \hline
    Model training time (s)  &167.8&107.6&146.1 \\
    \hline
    Prediction time ($\upmu$s) &22 &21 & 21 \\
    \hline
    Re-simulation RMSE (\%) &12.9& 4.1&2.7 \\
    \hline
    \end{tabular}
    \vspace{-6 pt}
\end{table}

Next, in the global optimization phase, the algorithm iteratively explores the continuous latent space $\mathcal{Z}$ to find the optimal design. During each iteration, the latent variables are input to the forward surrogate model of the VAE decoder. This rapidly predicts the corresponding EM responses, bypassing the need for expensive EM simulations. 

Concurrently, to account for physical realizability under given process constraints, the generated responses are first converted into compact physical features by the physics-aware parametric model, and the MDN then estimates the conditional distribution of feasible physical dimensions. A joint fitness function is then constructed. It not only measures the deviation between predicted responses and target metrics but also applies a physical penalty if the inferred geometric parameters exceed permissible boundaries or show low confidence probabilities. Guided by this combined fitness score, the PSO algorithm systematically updates the latent variables until convergence is reached.

Finally, the optimized latent variables are mapped back to the geometric space to extract the physical layout parameters. Subsequently, an EM simulation is conducted to verify the actual performance of the generated structure. If this verification fails, the latent space optimization is either restarted or subjected to further fine-tuning. Otherwise, the verified design is adopted as the initial layout level solution.

\subsection{Task-Dependent circuit level post optimization}
\label{sec:HFORD_Post}
As described in the HFORD architecture, the HFORD core maps unified device level requirements to a layout level seed. For complex circuit level tasks, this seed is not necessarily the final design but provides a physically feasible starting point for subsequent optimization, where the objective is defined by network level responses rather than isolated device level characteristics.

Let $\mathbf{x}_0$ denote the layout seed generated by the HFORD core. Starting from $\mathbf{x}_0$, the post optimization adjusts the passive layout parameters $\mathbf{x}$ and task-specific auxiliary variables $\mathbf{a}$ according to the final circuit level specifications. The network level response evaluated by field-circuit co-simulation is denoted as $\mathbf{c}_{\mathrm{sim}}(\mathbf{x},\mathbf{a})$. The post optimization is formulated as
\begin{equation}
\begin{aligned}
\min_{\mathbf{x},\,\mathbf{a}} \quad 
\mathcal{J}_{c}(\mathbf{x},\mathbf{a})=&\;\mathcal{L}_{c}\left(\mathbf{c}_{\mathrm{sim}}(\mathbf{x},\mathbf{a}),\mathbf{c}^{\star}\right)  \\
&+\lambda\left\|\mathbf{x}-\mathbf{x}_{0}\right\|_{2}^{2}+\alpha\Phi_{\mathrm{DRC}}(\mathbf{x}),
\end{aligned}
\label{eq:circuit_level_postopt}
\end{equation}
where $\mathbf{c}^{\star}$ denotes the target circuit level specification, and $\mathcal{L}_{c}(\cdot)$ measures the mismatch between the simulated and desired network level responses. The second term keeps the search close to the generated seed within a physically meaningful neighborhood, while $\Phi_{\mathrm{DRC}}(\mathbf{x})$ penalizes process and layout violations. The weights $\lambda$ and $\alpha$ balance the circuit objective, seed preservation, and manufacturability.
\begin{figure*}[!t]
\centering
\subfigure[][]{%
\label{Inductor}%
\includegraphics[height=1.82in]{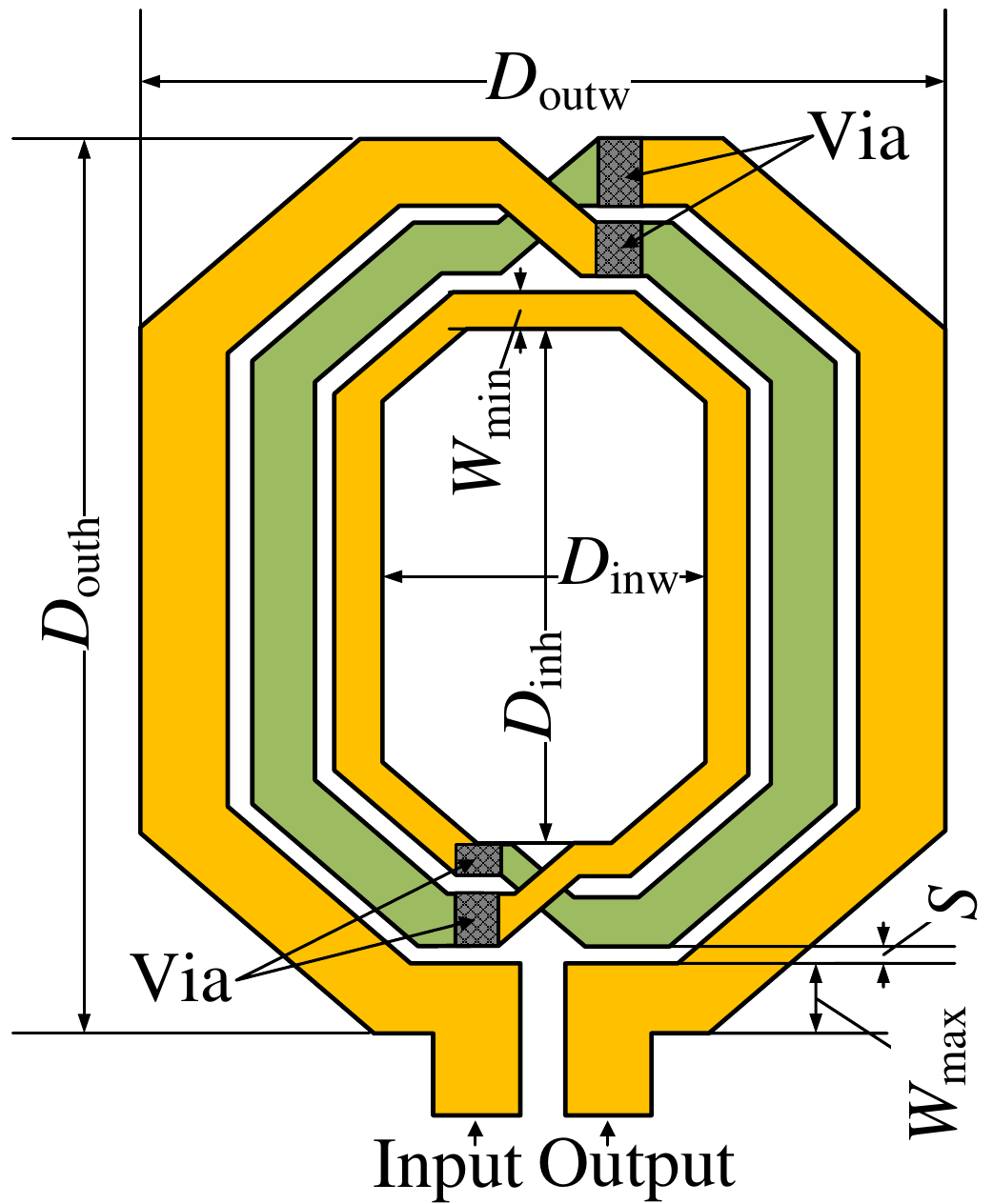}}
\hspace{10 pt}%
\subfigure[][]{%
\label{fig_L_Qmax}%
\includegraphics[height=1.82in]{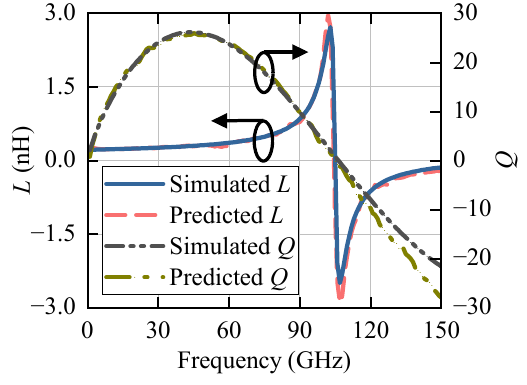}}
\hspace{10 pt}%
\subfigure[][]{%
\label{fig_Q_Qmax}%
\includegraphics[height=1.82in]{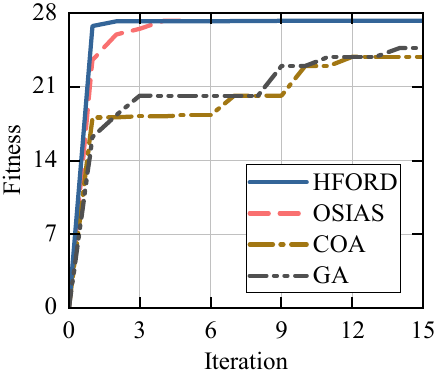}}
\caption{\subref{Inductor} The structure of a compressed symmetrical inductor and its geometrical parameters; \subref{fig_L_Qmax} comparison between predicted and simulated $L$, $Q$ responses; \subref{fig_Q_Qmax} convergence curves of different optimization methods.}%
\label{fig_Case_Qmax}
\vspace{-6 pt}
\end{figure*}
\begin{table*}
    \centering
    \caption{Comparisons of the Proposed HFORD with the GA, the COA and the OSIAS when Maximizing the Quality Factor}
    \label{Example1_tab}
    \begin{tabular}{|c|cccccc|c|c|c|c|}
    \hline
    Algorithm &$W_{\rm max}$ ($\upmu$m)&$S$ ($\upmu$m)&$W_{\rm min}$ ($\upmu$m)&$N$&$D_{\rm inw}$ ($\upmu$m)&$\gamma$&$L_{\rm 50GHz}$ (nH)&$Q_{\rm 50GHz}$&$Q_{\rm 100GHz}$&$t$ (h)\\
    \hline
    GA   &7.82&2.06&3.07&2&49.07&0.81 &0.31&24.88&1.89&29.26\\
    \hline
    COA~\cite{pierezan2018coyote}  &6.61&2.99&5.38&2&42.15 &0.99&0.30&27.25&5.78&15.74 \\
    \hline
    OSIAS~\cite{wei2023highly}  &6.76&2.71&5.26&2&48.26&0.75 &0.29&27.67&6.84&0.82 \\
    \hline
    HFORD &6.12&2.99&6.11&2&42.92 &0.99&0.29&27.54&8.18&0.03 \\
    \hline
    \end{tabular}
    \vspace{-6 pt}
\end{table*}

The form of $\mathcal{L}_{c}$ depends on the application and may include impedance matching, insertion loss, bandwidth, gain, phase balance, or in-band ripple. Therefore, this stage is kept as a task-dependent interface that can use different objective functions, auxiliary variables, and simulation flows. It avoids an unconstrained global search after seed generation and preserves the EM consistency inherited from the HFORD core. This separation allows the HFORD core to remain reusable across passive synthesis tasks without being tied to a specific matching theory, compensation strategy, or circuit topology.

\section{Design Examples}
\label{Sec_Results}
In this section, two mmWave on-chip inductive-element examples are used to validate the proposed HFORD method. The inductor example verifies device level synthesis, while the transformer example covers both device level performance optimization and circuit level impedance matching.

\subsection{Application to Millimeter-Wave On-chip Inductor}
Conventional on-chip inductor design relies on designer experience, surrogate modeling, and repeated EM verification because the geometric parameters are strongly coupled with $L$, $Q$, and SRF constraints. This example evaluates the proposed HFORD core for device level synthesis, where inductive element requirements are directly mapped to physical layouts. For a given target, the random forest selector first identifies a suitable inductor type, and the corresponding VAE and MDN models are then loaded for latent space layout generation under fabrication and DRC constraints.
\begin{table}[tbp]
    \begin{center}
    \caption{Ranges of Values for Samples}
    \label{Ranges_Samples}
    \begin{tabular}{| l | c |}
    \hline
    Parameter & Value range\\
    \hline
    Line width ($\rm \upmu m$) & $1.5\sim10$\\
    \hline
    Line space ($\rm \upmu m$)& $1.5\sim5$\\
    \hline
    Maximum area ($\rm \upmu m^2$)& 100$\times$100\\
    \hline
    Number of turns &$1\sim5$ \\
    \hline
    Deformation ratio & $0.1\sim3.0$\\
    \hline
    \end{tabular}
    \end{center}
    \vspace{-6 pt}
\end{table}

In greater detail, the inductors are fabricated in a 40-nm RF CMOS process with 3.4-$\upmu$m and 0.9-$\upmu$m metal layers, and the algorithm runs on a server with two AMD EPYC 7F52 processors and 512 GB of memory. A total of 5,000 inductor samples are used for training, and their geometric ranges are listed in Table~\ref{Ranges_Samples}. For the EMX simulations, the scanning range is set to [$0, 200$] GHz with a step size of 1 GHz, while the edge mesh is configured to 0.1 $\upmu$m, guaranteeing accuracy and efficiency.

The inverse model is implemented in PyTorch. After EM simulation, the values $L$ and $Q$ from 1 to 200 GHz are extracted. The network takes the frequency and the corresponding $L/Q$ values as inputs and outputs six design variables. The dataset is divided into training and test sets in an 8:2 ratio. A two-layer network with 200 nodes per layer and Tanh activation is trained for 1,000 iterations, with the number of Gaussian mixtures set to $m=5$.

The quality factor is one of the most critical parameters of an inductor and is closely related to the loss. The higher $Q$ indicates the better performance of inductor. For example, when designing an octagonal symmetrical deformable spiral inductor with $F_{\rm target}=50\ \rm GHz$, $L_{\rm target}=0.3\ \rm nH$, and ${\rm SRF}>100\ \rm GHz$, to maximize $Q$, the problem that must be solved is
\begin{equation}
    \begin{split}
    \label{MaxQ_eqt}
        &\max(Q_{\rm 50GHz}) \\
        &\begin{array}{l@{\quad}l@{}r@{\quad}r}
        \rm {s.t.}&\rm {geometrical\ constraints},\\
            &\frac{|L_{\rm WF}- L_{\rm WF\pm 5GHz}|}{L_{\rm WF}}<0.05,\\
            &\frac{|L_{\rm 50GHz}-L_{\rm target}|}{L_{\rm target}}<0.05,\\
            &{\rm SRF}>100\ \rm GHz,\\
        \end{array}
    \end{split}
\end{equation}
where $L_{\rm 50GHz}$ and $Q_{\rm 50GHz}$ are the predicted values used during the optimization of the latent space, and $L_{\rm target}$ is the input. 

For this objective, the automatic selector chooses the octagonal symmetrical inductor. The generated layout satisfies the physical dimensions and DRC constraints. The predicted responses are compared with the EM simulation results in Fig.~\ref{fig_Case_Qmax}, and the detailed metrics and optimization time are summarized in Table~\ref{Example1_tab}. Compared with previous optimization schemes under the same targets, HFORD achieves comparable performance with a much shorter optimization time.
\begin{figure*}[!t]
\centering
\subfigure[][]{%
    \label{transformer_2}%
    \includegraphics[height=0.205\textwidth]{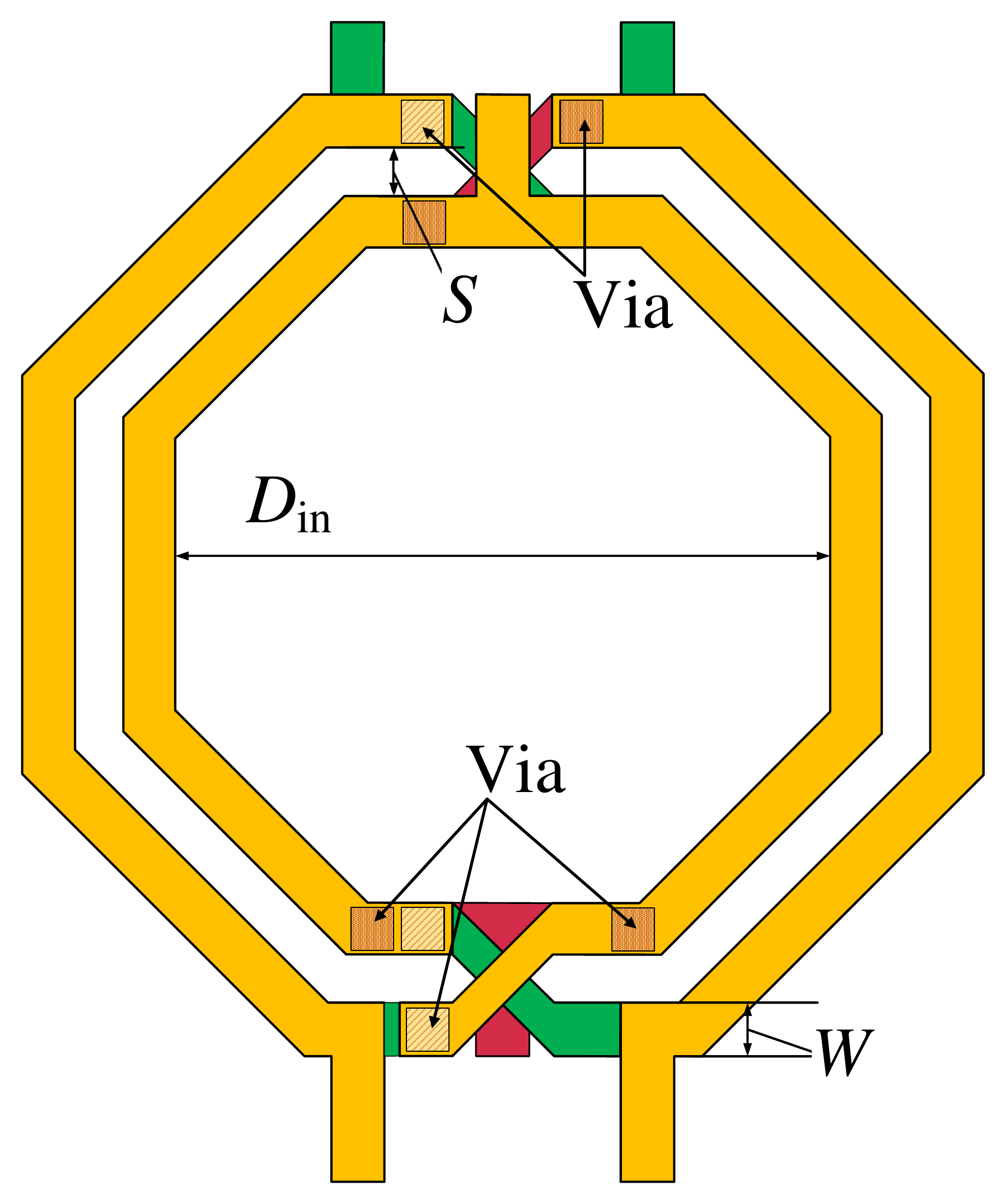}}
\hfill 
\subfigure[][]{%
    \label{G_max_Lp}%
    \includegraphics[width=0.235\textwidth]{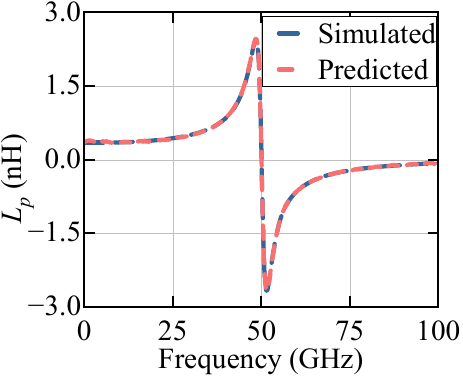}}
\hfill 
\subfigure[][]{%
    \label{G_max_Qp}%
    \includegraphics[width=0.235\textwidth]{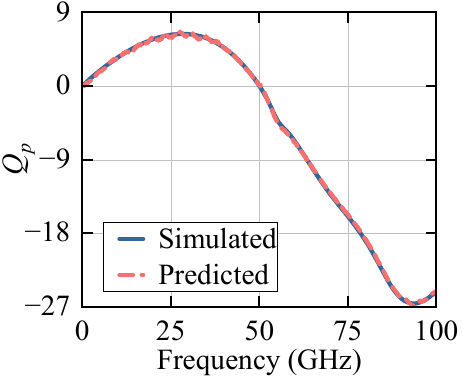}}
\hfill
\subfigure[][]{%
    \label{G_max_Ls}%
    \includegraphics[width=0.235\textwidth]{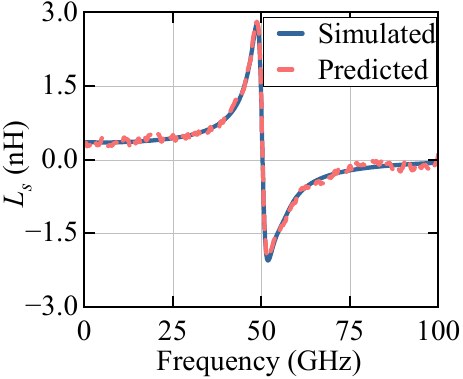}}
    
\vspace{6pt} 

\subfigure[][]{%
    \label{G_max_Qs}%
    \includegraphics[width=0.235\textwidth]{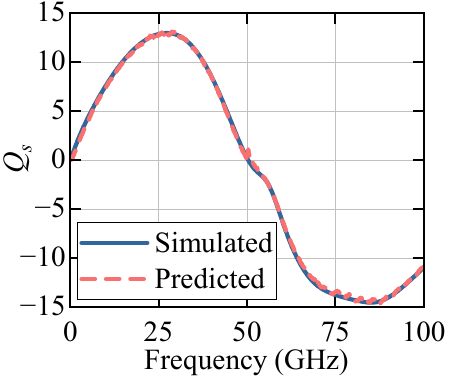}}
\hfill
\subfigure[][]{%
    \label{G_max_K}%
    \includegraphics[width=0.235\textwidth]{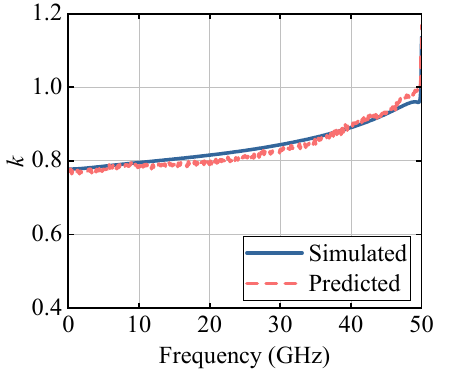}}
\hfill
\subfigure[][]{%
    \label{G_max_IL}
    \includegraphics[width=0.235\textwidth]{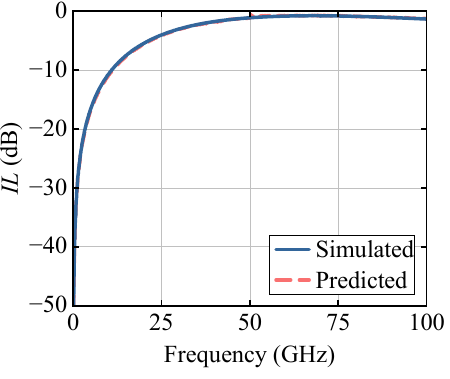}}
\hfill
\subfigure[][]{%
    \label{G_max_Gmax}
    \includegraphics[width=0.235\textwidth]{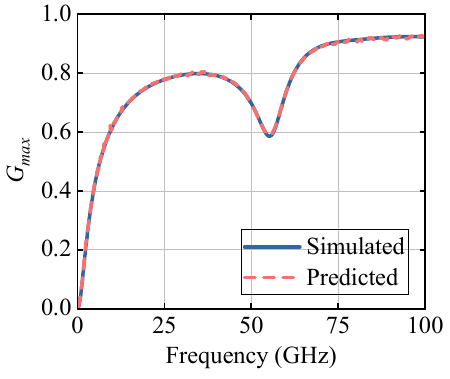}}

\caption{Performance prediction results of the on-chip transformer. 
\subref{transformer_2} Symmetrically overlapped octagonal transformer;
\subref{G_max_Lp} $L_p$; 
\subref{G_max_Qp} $Q_p$; 
\subref{G_max_Ls} $L_s$; 
\subref{G_max_Qs} $Q_s$; 
\subref{G_max_K} $k$; 
\subref{G_max_IL} $\mathrm{IL}$; 
\subref{G_max_Gmax} $G_{\rm max}$.}
\label{fig_Gmax}
\vspace{-6 pt}
\end{figure*}
\begin{table*}
    \centering
    \caption{Comparisons of the Proposed HFORD with the GA, the COA and the MLAO when Maximizing Gain}
    \label{Example2_tab}
    \begin{tabular}{|c|cccc|c|c|c|c|c|}
    \hline
    Algorithm &$W$ ($\upmu$m)&$S$ ($\upmu$m)&$N$&$D_{\rm inw}$ ($\upmu$m)&$k$ &$Q_{\rm 28GHz}$&$G_{\rm max}$&$t$ (h)\\
    \hline
    GA   &3.68&1.54&1&121.78 &0.70&11.68&0.79&42.34\\
    \hline
    COA~\cite{pierezan2018coyote}  &3.94&2.06&1&139.82&0.69&11.97&0.70&27.63 \\
    \hline
    MLAO~\cite{Wu2024} &2.46&1.65&1&143.01 &0.70&11.74&0.80&9.12 \\
    \hline
    HFORD &2.39&1.51&1&141.22&0.70&11.74&0.79&0.04 \\
    \hline
    \end{tabular}
    \vspace{-6 pt}
\end{table*}

\subsection{Application to Millimeter-Wave On-chip Transformer}
Compared with on-chip inductors, on-chip transformers have stronger coupling effects, more degrees of design freedom, and more response quantities, which make their modeling and inverse design more challenging.  To reduce the response dimensionality, the physics-based extraction described in Section~\ref{Sec_Physics_Aware} is used to maintain compactness and accuracy while preserving the dominant coupling and resonance characteristics.
\begin{table}[!t]
    \begin{center}
    \caption{Ranges of Values for Samples}
    \label{Ranges_Samples_trans}
    \begin{tabular}{| l | c |}
    \hline
    Parameter & Value range\\
    \hline
    Line width ($\rm \upmu m$) & $2.2 \sim 10$\\
    \hline
    Line space ($\rm \upmu m$)& $1.5 \sim 10$\\
    \hline
    Maximum area ($\rm \upmu m^2$)& 200$\times$200\\
    \hline
    Number of turns & $1\sim3$\\
    \hline
    Deformation ratio & $0.1\sim1.0$\\
    \hline
    Center offset & $0.1\sim0.5$\\
    \hline
    \end{tabular}
    \end{center}
    \vspace{-6pt}
\end{table}

The transformers are fabricated in the same 40-nm RF CMOS process and evaluated using the same computing platform. A total of 10,000 transformer samples are used for training, and their geometric ranges are listed in Table~\ref{Ranges_Samples_trans}. EMX simulations are performed from 0 to 100 GHz with a 0.1 GHz step, while the edge mesh is set to 0.1~$\upmu$m.
\begin{figure*}[!t]
\centering
\subfigure[][]{%
    \label{transformer_1}%
    \includegraphics[width=0.29\textwidth]{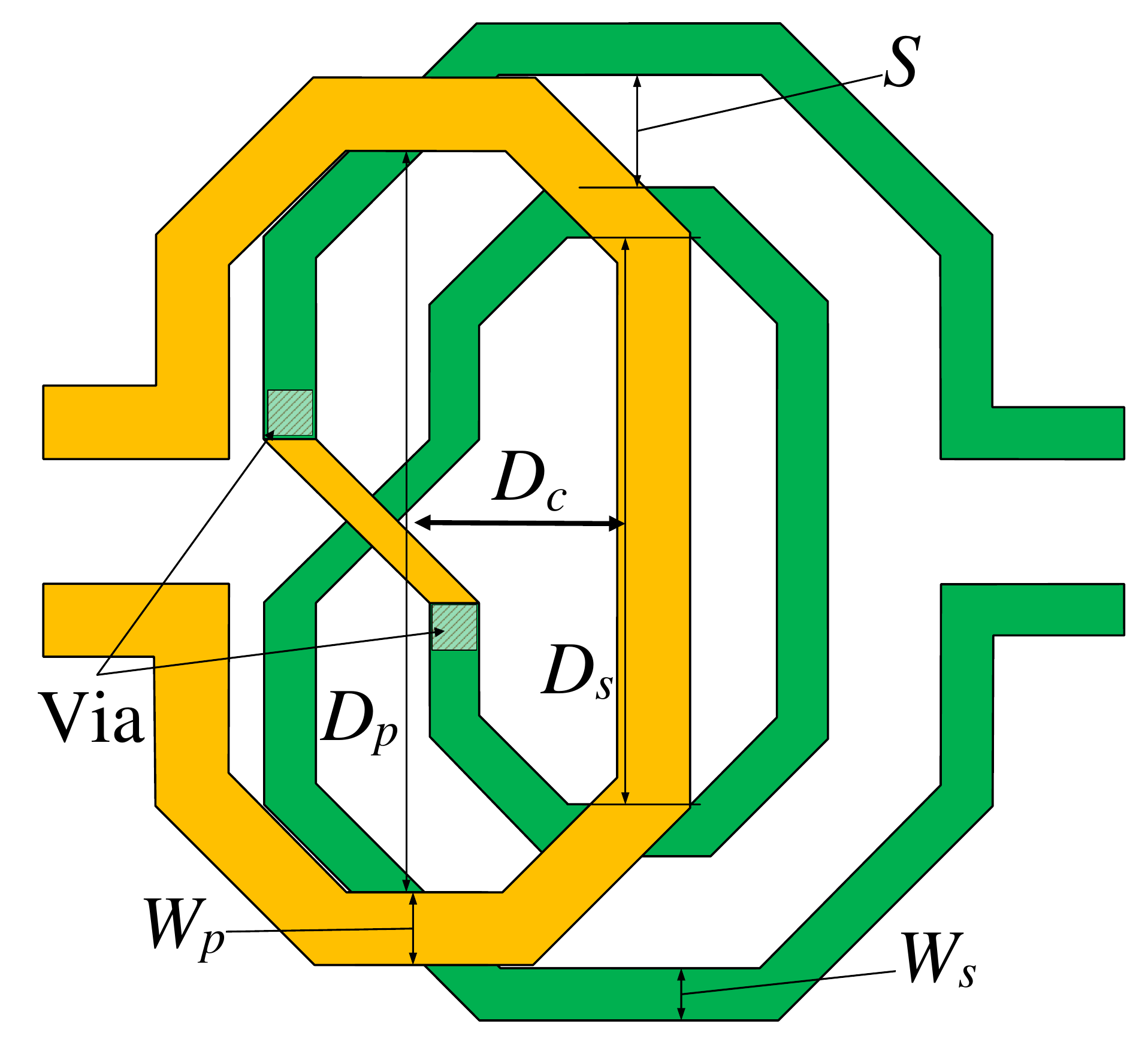}}
\hfill
\subfigure[][]{%
    \label{BM_Z}
    \includegraphics[width=0.31\textwidth]{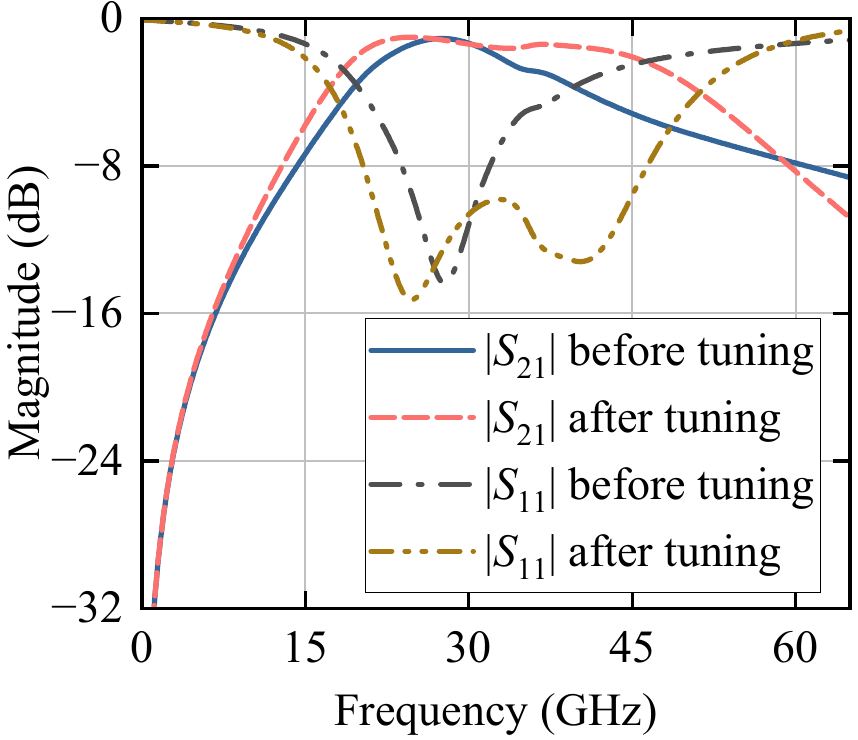}}
\hfill
\subfigure[][]{%
    \label{BM_S}
    \includegraphics[width=0.335\textwidth]{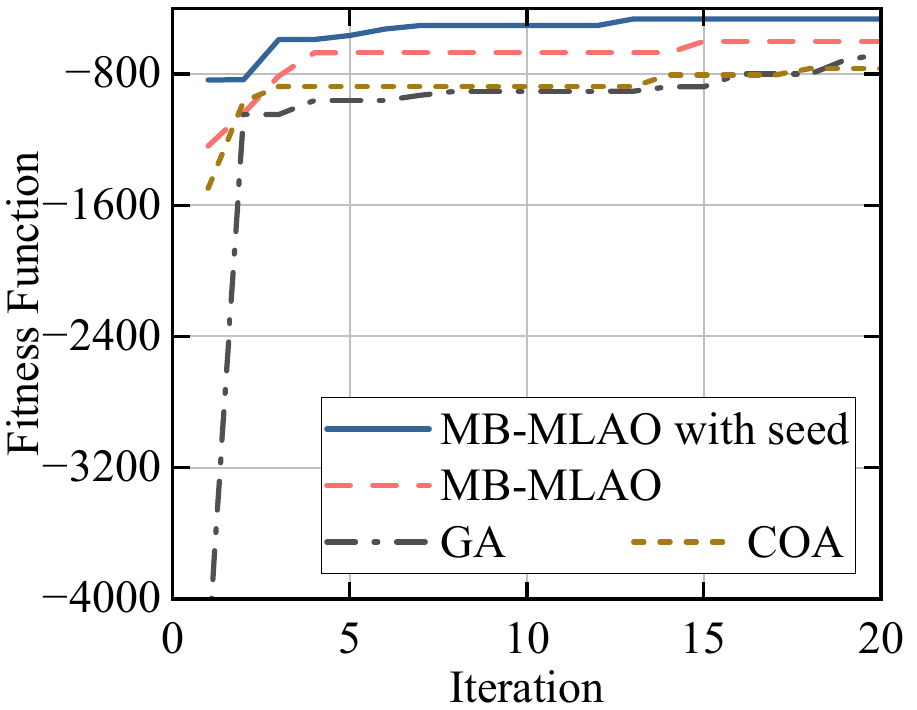}}
\caption{Impedance-matching design results.
\subref{transformer_1} 1:2 Symmetrically overlapped transformer;
\subref{BM_Z} $S$-parameter before and after tuning;
\subref{BM_S} Fitness convergence comparison of different optimization methods.}
\label{fig_BW}
\vspace{-6 pt}
\end{figure*}

Based on the simulated Z-parameters and S-parameters, $L_p$, $L_s$, $Q_p$, $Q_s$, $M$, and $Re(Z_{21})$ are extracted. During training, the target frequency and the corresponding EM response matrix are used as input, while the output consists of six transformer geometry parameters. The dataset is divided into training and test sets in an 8:2 ratio. The model uses two hidden layers with 200 nodes per layer and Tanh activation and is trained for 2,000 iterations with $m=5$.
\begin{figure}[htpb]
\centering
\subfigure[][]{%
\label{RLC}%
\includegraphics[width=1.65in]{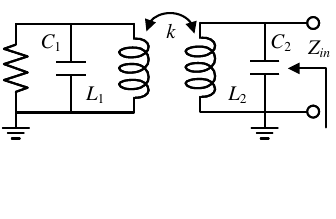}}
\hspace{-1pt}%
\subfigure[][]{%
\label{BW_Smith}%
\includegraphics[width=1.65in]{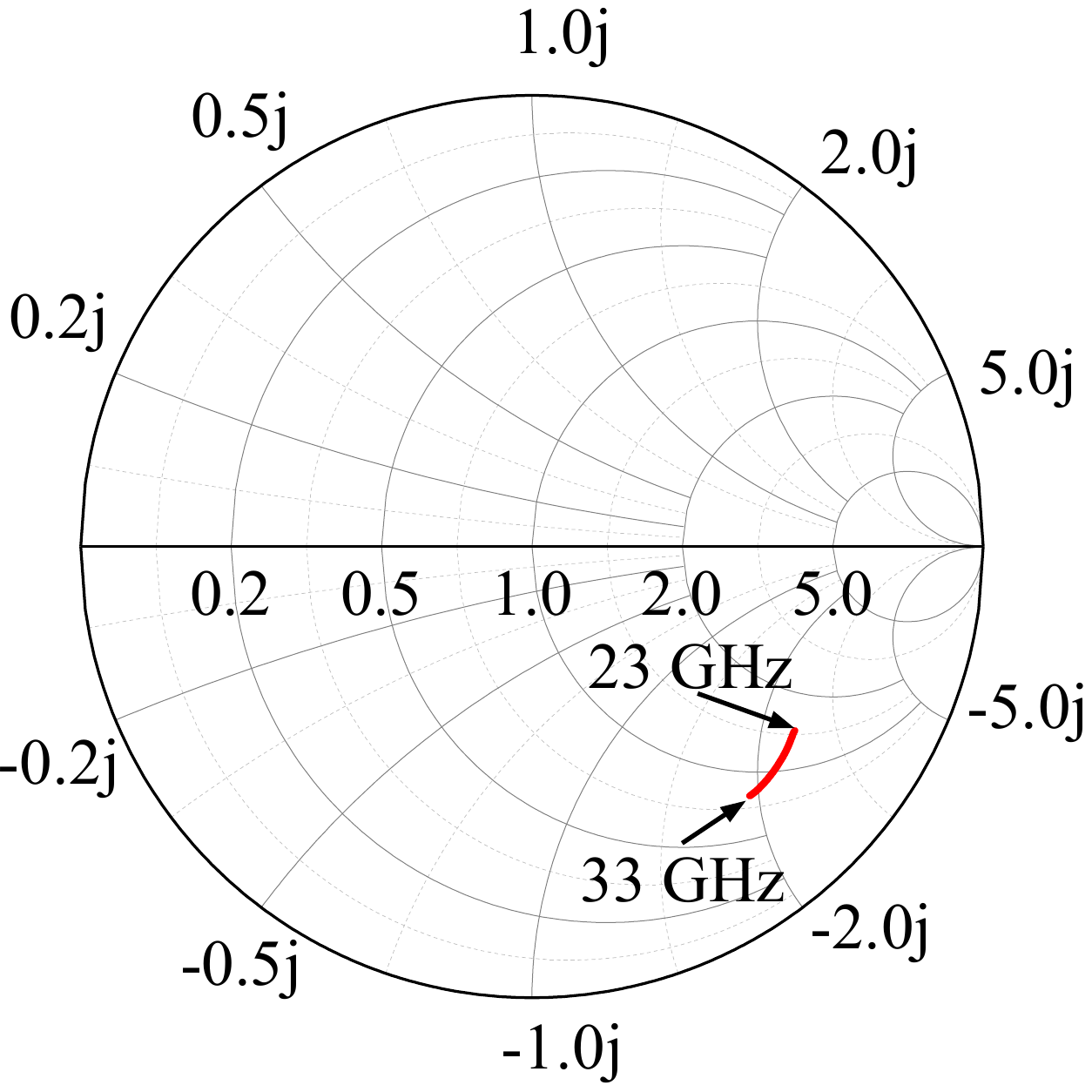}}
\hspace{-1pt}%
\caption{Broadband input matching using the coupled-resonator model. \subref{RLC} Equivalent circuit. \subref{BW_Smith} Input impedance trajectory on the Smith chart.}
\label{fig:matching_model}
\vspace{-6 pt}
\end{figure}

The transformer library covers quadrilateral and octagonal geometries, interleaved and stacked configurations, and both equal-turn and unequal-turn designs. The quadrilateral interleaved and quadrilateral stacked transformers are selected as base templates, and parameter-transfer training is applied to the remaining transformer types.

Two transformer design scenarios are considered. The first evaluates device level performance optimization by maximizing $G_{\max}$ of a transformer. The second evaluates circuit level impedance matching, where the HFORD-generated transformer seed is refined through task-dependent post optimization. The overall workflow is shown in Fig.~\ref{HFORD}.

\subsubsection{Available Gain Maximization}
\ 
\newline
\indent In mmWave receiver front-ends, the transformer strongly affects signal transmission efficiency. For this application, a differentially stacked topology is selected to exploit the vertical magnetic coupling between the top and lower metal layers. Although this structure may reduce SRF due to the increased inter-winding capacitance, the optimization flow balances this trade-off to maximize $G_{\rm max}$ at the operating frequency.

The objective is to maximize $G_{\rm max}$ at 28 GHz under physical design rules and bandwidth constraints, which is formulated as
\begin{equation}
    \begin{split}
    \label{Gmax_eqt}
        &\max_{\mathbf{x}} \quad G_{\rm max}(\mathbf{x}) \big|_{f=28\ \text{GHz}} \\
        &\begin{array}{l@{\quad}l@{}r@{\quad}r}
        \rm {s.t.}&\rm {geometrical\ constraints},\\
            &\frac{|L_{\rm WF}- L_{\rm WF\pm 5GHz}|}{L_{\rm WF}}<0.05,\\
            &\frac{|L_{\rm 28GHz}-L_{\rm target}|}{L_{\rm target}}<0.05,\\
            &\text{SRF} \ge 45\; \text{GHz}, \quad \mathbf{x} \in \mathit{\Omega}_{\rm drc}\\
        \end{array}
    \end{split}
\end{equation}

As shown in Fig.~\ref{fig_Gmax} and Table~\ref{Example2_tab}, the generated design satisfies the specifications and shows good agreement between the predicted and simulated responses. Compared with conventional optimization methods, HFORD achieves similar performance with a much shorter optimization time, demonstrating its effectiveness for device level transformer gain optimization under physical and DRC constraints.

\subsubsection{Impedance Matching}
\ 
\newline
\indent For the input matching network of a power amplifier, broadband impedance matching and in-band transmission flatness are required to drive the subsequent circuit stage. A stacked 1:2 transformer topology is adopted because it provides high $k$ and additional geometrical freedom for broadband impedance transformation.

The synthesis follows the task-dependent post optimization flow in Section~\ref{sec:HFORD_Post}. As illustrated in Fig.~\ref{fig:matching_model}, analytical optimization based on magnetic coupling resonance is first used to estimate the initial lumped parameters, including $L$, $k$, and compensation capacitance $C$. Based on these electrical targets, the HFORD core generates an initial transformer seed through latent space optimization. Since layout parasitics cannot be fully captured by fixed $L$ and $k$, the seed is further refined by field-circuit co-simulation, including transformer geometry adjustments and compensation capacitor tuning for $C_p$ and $C_s$.

\begin{table*}[!t]
\centering
\caption{Performance Comparisons of the Proposed HFORD Method and Those in References}
\label{tab:qualitative_comparison}
\begingroup
\footnotesize
\setlength{\tabcolsep}{1.8pt}
\renewcommand{\arraystretch}{1.22}
\newcolumntype{L}[1]{>{\raggedright\arraybackslash}m{#1}}
\newcolumntype{C}[1]{>{\centering\arraybackslash}m{#1}}

\begin{tabular}{|C{1.55cm}|C{1.75cm}|C{3.05cm}|C{3.4cm}|C{2.10cm}|C{2.45cm}|C{2.35cm}|}
\hline
Reference &
Application &
Auto Topology Selection &
Reported Model Error &
Synthesis Time &
Response Dimension &
Illusion Mitigation \\
\hline

\multirow{2}{1.55cm}{\centering This work}
& Inductor
& Yes
& MAE $<$ 0.04
& 2 min
& 6-D
& Yes \\
\cline{2-7}

& Transformer
& Yes
& MAE $<$ 0.06
& 3 min
& 40-D
& Yes \\
\hline


\cite{chae2024pulserf}
& Transformer
& N.A.
& MAE $<$ 0.06
& 3--9 min
& 12$\times$300
& Partial \\
\hline

\cite{he2025motif}
& Transformer
& No
& MAE $<$ 0.01
& $<$3 min
& 2400-D
& Partial \\
\hline

\cite{lee2022transformer}
& Transformer
& No
& MSE $<$ $3.03{\times}10^{-5}$
& 10 min
& 100$\times$8
& No \\
\hline
\end{tabular}

\vspace{1mm}
\parbox{0.98\textwidth}{\footnotesize
N.A. denotes not applicable because the method is template-free rather than topology-selection based. Model Dimension denotes the reported neural-network response dimension. The model errors are reported following the metrics used in the corresponding references. Since different works evaluate different response quantities and use different error definitions, these values are intended to indicate the reported modeling accuracy within each work rather than a strictly normalized cross-paper comparison.}
\endgroup
\vspace{-6pt}
\end{table*}

The objective is to match a $50$ $\Omega$ source to the input impedance of the subsequent encapsulated circuit stage, which is characterized by an imported .SNP file shown in Fig.~\ref{fig:matching_model}. Let $Z_S=50 \; \Omega$ denote the source impedance and $Z_L(f)$ denote the load impedance extracted from the .SNP file. Over the target band $23~\text{GHz}\le f\le 33~\text{GHz}$, the design task is formulated as
\begin{equation}
\begin{aligned}
\max_{\mathbf{x}}\quad & 
   \overline{|S_{21}|}\;-\;\lambda\,\Delta|S_{21}| \\[2pt]
\text{s.t.}\quad 
   & |S_{11}|\le -7\,\text{dB}, \\
   & |S_{21}|\ge -2\,\text{dB}, \\
   & \mathbf{x}\in\Omega_{\text{drc}},
\end{aligned}
\label{eq:matching_opt}
\end{equation}
where $\overline{|S_{21}|}$ denotes the in-band mean of $|S_{21}|$ in the target band, and $\Delta|S_{21}|$ is defined as the difference between the maximum and minimum values of $|S_{21}|$ within the band. The coefficient $\lambda>0$ is a weighting factor that trades off mean transmission against flatness, and $\Omega_{\text{drc}}$ denotes the feasible set imposed by the geometric constraints.

The optimization results are shown in Fig.~\ref{fig_BW}. Compared with the baselines, the proposed flow provides a more effective initialization for circuit level refinement, leading to improved agreement between the predicted seed behavior and the final field-circuit co-simulation results. A qualitative comparison with recent reverse design methods is summarized in Table~\ref{tab:qualitative_comparison}. The proposed HFORD delivers exceptional accuracy while requiring the least computational time. Additionally, it successfully eliminates the adverse illusory effects that have long plagued existing synthesis approaches. 

\section{Conclusion}
\label{Sec_Conc}
A hybrid target-to-layout synthesis method for mmWave on-chip inductive elements has been presented. A hierarchical flow unifies device- and circuit-level specifications, utilizing a reusable HFORD core for layout seed generation and post-optimization for circuit refinement. To ensure highly efficient layout generation, the framework integrates performance-aware sparse-fitting sampling and physics-informed parametric modeling alongside random-forest, VAE, MDN, and PSO algorithms. Validation on mmWave inductors and transformers confirms that the proposed method successfully generates DRC-compliant layouts and reduces the design cycle from hours to minutes compared to conventional forward optimization methods. Validated in a 40-nm RF CMOS process, the modular HFORD framework will be extended to different PDKs, technology nodes, and broader passive synthesis tasks in future work.
 
\section*{Acknowledgment}
Generative artificial intelligence (AI) tools were utilized in the preparation of this manuscript. Specifically, OpenAI ChatGPT and Google Gemini were used for minor grammar corrections, improving sentence structure, and refining the academic tone during the proofreading and editing phases of manuscript development. The authors confirm that all substantive intellectual content, research design, data analysis, and conclusions remain the original work of the authors, who take full responsibility for the content of this publication.  

\balance
\bibliographystyle{IEEEtran}
\bibliography{ref}

@article{Cui2003,
  title={Robust design of absorbers using genetic algorithms and the finite element-boundary integral method},
  author={Cui, S. and Weile, D. S.},
  journal={IEEE Trans. Antennas Propagat.},
  year={2003},
  volume={51},
  number={12},
  pages={3249--3258},
  doi={10.1109/TAP.2003.820971}
}

@book{meyer2000design,
  title={Design, simulation and applications of inductors and transformers for {Si RF ICs}},
  author={Meyer, Robert G},
  year={2000},
  publisher={Springer}
}

@article{Ozgun2003,
  title={Design of dual-frequency probe-fed microstrip antennas with genetic optimization algorithm},
  author={Ozgun, O. and Mutlu, S. and Aksun, M. I. and Alatan, L.},
  journal={IEEE Trans. Antennas Propagat.},
  year={2003},
  volume={51},
  number={8},
  pages={1947--1954},
  doi={10.1109/TAP.2003.814732}
}

@article{Ashinnmanesh2008,
  title={Design of a single-feed dual-band dual-polarized printed microstrip antenna using a Boolean particle swarm optimization},
  author={Ashinnmanesh, F. and Dorandi, A. and Shahabadi, M.},
  journal={IEEE Trans. Antennas Propagat.},
  year={2008},
  volume={56},
  number={7},
  pages={1845--1852},
  doi={10.1109/TAP.2008.924684}
}

@ARTICLE{Wu2024,
  author={Wu, Qi and Chen, Weiqi and Yu, Chen and Wang, Haiming and Hong, Wei},
  journal={IEEE Trans. Antennas Propagat.}, 
  title={Machine-Learning-Assisted Optimization for Antenna Geometry Design}, 
  year={2024},
  volume={72},
  number={3},
  pages={2083-2095},
  doi={10.1109/TAP.2023.3346493}}

@article{Nayeri2013,
  title={Design of single-feed reflectarray antennas with asymmetric multiple beams using the particle swarm optimization method},
  author={Nayeri, P. and Yang, F. and Elsherbeni, A. Z.},
  journal={IEEE Trans. Antennas Propagat.},
  year={2013},
  volume={61},
  number={9},
  pages={4598--4605},
  doi={10.1109/TAP.2013.2268243}
}

@article{Koziel2021,
  title={Machine-learning-powered {EM}-based framework for efficient and reliable design of low scattering metasurfaces},
  author={Koziel, S. and Abdullah, M.},
  journal={IEEE Trans. Microwave Theory Techn.},
  year={2021},
  volume={69},
  number={4},
  pages={2028--2041},
  doi={10.1109/TMTT.2021.3061128}
}

@article{Prado2019,
  title={Wideband shaped-beam reflectarray design using support vector regression analysis},
  author={Prado, D. R. and L{\'o}pez-Fern{\'a}ndez, J. A. and Arrebola, M. and Rodr{\'i}guez-Pi{\~n}o, M. and Goussetis, G.},
  journal={IEEE Antennas Wireless Propag. Lett.},
  year={2019},
  volume={18},
  number={11},
  pages={2287--2291},
  doi={10.1109/LAWP.2019.2932902}
}

@article{Jacobs2013,
  title={Computationally efficient multi-fidelity Bayesian support vector regression modeling of planar antenna input characteristics},
  author={Jacobs, J. P. and Koziel, S. and Ogurtsov, S.},
  journal={IEEE Trans. Antennas Propagat.},
  year={2013},
  volume={61},
  number={2},
  pages={980--984},
  doi={10.1109/TAP.2012.2220513}
}

@article{wei2023highly,
  author  = {Wei, J. and others},
  journal = {IEEE Trans. Comput.-Aided Des. Integr. Circuits Syst.},
  title   = {Highly efficient automatic synthesis of a millimeter-wave on-chip deformable spiral inductor using a hybrid knowledge-guided and data-driven technique},
  year    = {2023},
  volume  = {42},
  number  = {12},
  pages   = {4413--4422},
  doi     = {10.1109/TCAD.2023.3281389}
}

@article{He2024,
  title={Hybrid method of artificial neural network and simulated annealing algorithm for optimizing wideband patch antennas},
  author={He, Y. and Huang, J. and Li, W. and Zhang, L. and Wong, S.-W. and Chen, Z. N.},
  journal={IEEE Trans. Antennas Propagat.},
  year={2024},
  volume={72},
  number={1},
  pages={944--949},
  doi={10.1109/TAP.2023.3331249}
}

@article{Wu2020,
  title={Multistage collaborative machine learning and its application to antenna modeling and optimization},
  author={Wu, Q. and Wang, H. and Hong, W.},
  journal={IEEE Trans. Antennas Propagat.},
  year={2020},
  volume={68},
  number={5},
  pages={3397--3409},
  doi={10.1109/TAP.2019.2963570}
}

@article{Xiao2018,
  title={Dynamic adjustment kernel extreme learning machine for microwave component design},
  author={Xiao, L.-Y. and Shao, W. and Ding, X. and Wang, B.-Z.},
  journal={IEEE Trans. Microwave Theory Techn.},
  year={2018},
  volume={66},
  number={10},
  pages={4452--4461},
  doi={10.1109/TMTT.2018.2858787}
}

@article{Zhang2018,
  title={Space mapping approach to electromagnetic-centric multiphysics parametric modeling of microwave components},
  author={Zhang, W. and others},
  journal={IEEE Trans. Microwave Theory Techn.},
  year={2018},
  volume={66},
  number={7},
  pages={3169--3185},
  doi={10.1109/TMTT.2018.2831220}
}

@article{Prado20192,
  title={Support vector regression to accelerate design and cross-polar optimizations of shaped-beam reflectarray antennas},
  author={Prado, D. R. and L{\'o}pez-Fern{\'a}ndez, J. A. and Arrebola, M. and Goussetis, G.},
  journal={IEEE Trans. Antennas Propagat.},
  year={2019},
  volume={67},
  number={3},
  pages={1659--1668},
  doi={10.1109/TAP.2018.2880299}
}

@article{Zhou2023,
  title={Representation learning-driven fully automated framework for the inverse design of frequency-selective surfaces},
  author={Zhou, Z. and Wei, Z. J. and Ren, Y. and Yin, Y. and Pedersen, G. F. and Shen, M.},
  journal={IEEE Trans. Microwave Theory Techn.},
  year={2023},
  volume={71},
  number={6},
  pages={2409--2421},
  doi={10.1109/TMTT.2023.3250068}
}

@article{Zhang2018MNN,
  title={Multivalued neural network inverse modeling and applications to microwave filters},
  author={Zhang, C. and Jin, J. and Na, Q. and Zhang, M. and Yu, M.},
  journal={IEEE Trans. Microwave Theory Techn.},
  year={2018},
  volume={66},
  number={8},
  pages={3781--3797},
  doi={10.1109/TMTT.2018.2841889}
}

@article{Yuan.metasurfaces,
  title={An efficient artificial neural network model for inverse design of metasurfaces},
  author={Yuan, L. and Wang, X.-S. and Huang, H. and Wang, B.-Z.},
  journal={IEEE Antennas Wireless Propag. Lett.},
  year={2021},
  volume={20},
  number={6},
  pages={2013--2017},
  doi={10.1109/LAWP.2021.3069713}
}

@article{MCVAE,
  title={End-to-End Machine-Learning Framework for Electromagnetic Inverse Design: From Practical Constraints to Optimized Structures},
  author={Zhou, Z. and Wei, Z. H. and Ren, J. and Yin, Y. Z. and Li, J. and Chan, T.-T.},
  journal={IEEE Trans. Microwave Theory Techn.},
  year={2025},
  volume={73},
  number={11},
  pages={8690--8708},
  doi={10.1109/TMTT.2025.3583316}
}

@inproceedings{Kingma2014Auto,
    author = {D. P. Kingma and M. Welling},
    title = {Auto-encoding variational {Bayes}},
    booktitle = {2nd International Conference on Learning Representations, ICLR 2014, Banff, AB, Canada, April 14--16, 2014, Conference Track Proceedings},
    year = {2014}
}

@ARTICLE{Zhang2024,
  author={Zhang, Yuwei and Xu, Jinping},
  journal={IEEE Microw. Wireless Technol. Lett.}, 
  title={Inverse Design of Dual-Band Microstrip Filters Based on Generative Adversarial Network}, 
  year={2024},
  volume={34},
  number={1},
  pages={29-32},
  doi={10.1109/LMWT.2023.3329047}}

@book{Bishop2006,
  author    = {Bishop, C. M.},
  title     = {Pattern Recognition and Machine Learning},
  publisher = {Springer},
  address   = {New York, NY, USA},
  year      = {2006}
}

@article{breiman2001random,
  author  = {Breiman, Leo},
  title   = {Random Forests},
  journal = {Machine Learning},
  volume  = {45},
  number  = {1},
  pages   = {5--32},
  year    = {2001},
  doi     = {10.1023/A:1010933404324}
}

@inproceedings{pierezan2018coyote,
  author={Pierezan, Juliano and Dos Santos Coelho, Leandro},
  booktitle={2018 IEEE Congress on Evolutionary Computation (CEC)}, 
  title={Coyote Optimization Algorithm: A New Metaheuristic for Global Optimization Problems}, 
  year={2018},
  volume={},
  number={},
  pages={1--8},
  doi={10.1109/CEC.2018.8477769}}

@article{8132183,
  author={Rad, Ali Ajdari and Hasler, Martin and Jalili, Mahdi},
  journal={Logic Journal of the IGPL}, 
  title={Reservoir optimization in recurrent neural networks using properties of {Kronecker} product}, 
  year={2010},
  volume={18},
  number={5},
  pages={670--685},
  doi={10.1093/jigpal/jzp044}}

@inproceedings{NMMSO,
  author={Fieldsend, Jonathan E.},
  booktitle={2014 IEEE Congress on Evolutionary Computation (CEC)}, 
  title={Running Up Those Hills: Multi-modal search with the niching migratory multi-swarm optimiser}, 
  year={2014},
  volume={},
  number={},
  pages={2593--2600},
  doi={10.1109/CEC.2014.6900309}}

@article{McKay1979LHS,
  author    = {M. D. McKay and R. J. Beckman and W. J. Conover},
  title     = {A Comparison of Three Methods for Selecting Values of Input Variables in the Analysis of Output from a Computer Code},
  journal   = {Technometrics},
  volume    = {21},
  number    = {2},
  pages     = {239--245},
  year      = {1979},
  publisher = {Taylor \& Francis}
}

@book{Forrester2008Surrogate,
  author    = {A. I. J. Forrester and A. S\'{o}bester and A. J. Keane},
  title     = {Engineering Design via Surrogate Modelling: A Practical Guide},
  publisher = {John Wiley \& Sons},
  year      = {2008},
  address   = {Chichester, UK}
}

@inproceedings{he2025motif,
  author    = {Houbo He and Yizhou Xu and Lei Xia and Yaolong Hu and Fan Cai and Taiyun Chi},
  title     = {{MOTIF-RF}: Multi-Template On-Chip Transformer Synthesis Incorporating Frequency-Domain Self-Transfer Learning for {RFIC} Design Automation},
  booktitle = {Proceedings of the Asia and South Pacific Design Automation Conference (ASP-DAC)},
  year      = {2026},
  pages     = {1138--1144},
  doi       = {10.1109/ASP-DAC66049.2026.11420464}
}

@ARTICLE{Gupta2023,
  author={Gupta, Aggraj and Karahan, Emir Ali and Bhat, Chandan and Sengupta, Kaushik and Khankhoje, Uday K.},
  journal={IEEE Trans. Antennas Propagat.}, 
  title={Tandem Neural Network Based Design of Multiband Antennas}, 
  year={2023},
  volume={71},
  number={8},
  pages={6308-6317},
  doi={10.1109/TAP.2023.3276524}}

@article{karahan2024nat,
  author  = {Emir Ali Karahan and Zheng Liu and Aggraj Gupta and Zijian Shao and Jonathan Zhou and Uday Khankhoje and Kaushik Sengupta},
  title   = {Deep-Learning Enabled Generalized Inverse Design of Multi-Port Radio-Frequency and Sub-Terahertz Passives and Integrated Circuits},
  journal = {Nat. Commun.},
  volume  = {15},
  number  = {10734},
  year    = {2024},
  doi     = {10.1038/s41467-024-54178-1}
}

@article{lee2022transformer,
author  = {Dongyoon Lee and Gibeom Shin and Seunghoon Lee and Kyunghwan Kim and Tae-Hyun Oh and Ho-Jin Song},
title   = {Neural-Network-Based Automated Synthesis of Transformer Matching Circuits for {RF} Amplifier Design},
journal = {IEEE Trans. Microwave Theory Techn.},
volume  = {70},
number  = {11},
pages   = {4726--4739},
year    = {2022},
doi     = {10.1109/TMTT.2022.3199756}
}

@inproceedings{chae2024pulserf,
  author    = {Chae, Hyunsu and Yu, Hao and Li, Sensen and Pan, David Z.},
  title     = {{PulseRF}: Physics Augmented {ML} Modeling and Synthesis for High-Frequency {RFIC} Design},
  booktitle = {Proc. IEEE/ACM Int. Conf. Computer-Aided Design (ICCAD)},
  year      = {2024},
  pages     = {1--9}
}

\vfill

\end{document}